\documentclass{aa}  
\usepackage{natbib}
\usepackage{xcolor}
\usepackage{subcaption}
\usepackage[colorlinks=true, citecolor=magenta, linkcolor=red, urlcolor=magenta]{hyperref}
\bibpunct{(}{)}{;}{a}{}{,}
\usepackage{txfonts}
\usepackage{graphicx}
\usepackage{booktabs}
\usepackage{multirow}
\def\gaia{\textsl{Gaia}}
\newcommand{\photmoh}{[M/H]$_{\rm phot}$}
\usepackage[T1]{fontenc}
\usepackage[utf8]{inputenc}
\usepackage{CJKutf8}
\usepackage{listings}
\usepackage{amsmath}
\usepackage{comment}
\usepackage{tabularx}
\usepackage{booktabs}
\definecolor{mauve}{rgb}{0.88, 0.69, 1.0}
\definecolor{oldmauve}{rgb}{0.4, 0.19, 0.28}
\newcommand{\moh}{[M/H]$_{\rm phot}$}                     

\newcommand{\chisqimprove}{$\log_{10}(\chi^2_{\text{single}} / \chi^2_{\text{binary}})$}

\begin{document} 
\titlerunning{MSMS Binaries from Gaia BP/RP}

\authorrunning{Jiadong Li et al.}

   \title{Millions of Main-Sequence Binary Stars from Gaia BP/RP Spectra}


\author{Jiadong Li\inst{1}
       \and
       Hans-Walter Rix\inst{1}
       \and
       Yuan-Sen Ting\inst{2,3,1}
       \and
       Johanna M\"uller-Horn\inst{1}
       \and
       Kareem El-Badry\inst{4,1}
       \and
       Chao Liu\inst{5,6,7,8}
       \and
       Rhys Seeburger\inst{1}
       \and
       Gregory M. Green\inst{1}
       \and
       Xiangyu Zhang\inst{1}
       }

\institute{Max-Planck-Institut für Astronomie, Königstuhl 17, D-69117 Heidelberg, Germany\\
          \email{jdli@mpia.de}
         \and
         Department of Astronomy, The Ohio State University, Columbus, OH 43210, USA
         \and
         Center for Cosmology and AstroParticle Physics (CCAPP), The Ohio State University, Columbus, OH 43210, USA
         \and
         Department of Astronomy, California Institute of Technology, 1200 E. California Blvd., Pasadena, CA 91125, USA
         \and
         Key Lab of Space Astronomy and Technology, National Astronomical Observatories, Beijing, 100101, China
         \and
         Institute for Frontiers in Astronomy and Astrophysics of Beijing Normal University, Beijing, 100875, China
         \and
         University of Chinese Academy of Sciences, Beijing, 100049, China
         \and
         Zhejiang Lab, Hangzhou, 311121, China
         }
 
\abstract{
We present an extensive catalog of likely main-sequence binary stars systems (MSMS), derived from Gaia Data Release 3 BP/RP (XP) spectra through the comparison of single- and binary-star model fits. Leveraging the large sample of low-resolution Gaia XP spectra, we use a neural network to build a forward modeling for spectral luminosities of single-stars, as a function of stellar mass and photometric metallicity. Applying this model to XP spctra, we find that this enables the identification of binaries with mass ratios between 0.5 and 1.0 and flux ratios $>0.1$ as ``poor" fits to the data, either in spectral shape or flux normalization. From an initial sample of 35 million stars within 1 kpc, we identify 14 million possible binary candidates, and a high-confidence "golden sample" of 1 million binary systems. This large, homogeneous sample of SED-based binaries enables population studies of luminous MSMS binaries, and -- in conjunction with kinematic or astrometric probes -- permits to identify binaries with dark or dim companions, such as white dwarfs, neutron stars and black hole candidates, improving our understanding of compact object populations.
}
   
\maketitle
\section{Introduction}

Binary star systems constitute fundamental components of the stellar population, with half of solar-type stars residing in multiple systems \citep{raghavan2010, duchene2013, moe2019}. 
Binary systems provide an important context and set of constraints for most sub-fields of astrophysics \citep{rix2019}.
The statistical properties of binary populations—including the binary fraction, period distribution, mass ratio distribution, and eccentricity distribution—provide constraints on star formation theories and dynamical evolution models.
Moreover, these characteristics vary systematically with stellar properties such as stellar mass, age, and metallicity, creating distinctive patterns across different stellar populations and offering valuable insights into the physical mechanisms governing both the formation and subsequent evolution of stellar systems \citep{duchene2013, moe2019, el-badry2021}.

The periods of binary systems span a wide range in timescales (from minutes to Myr). 
Current stellar surveys such as \gaia, LAMOST, SDSS-IV/APOGEE and SDSS-V \citep{ Gaia2016, Liu2020, Majewski2017, Kollmeier2017} primarily identify binaries through spectroscopic \citep{el-badry2018, price-whelan2020, zhang2022a} or astrometric signatures of orbital motion \citep[e.g.][]{holl2023, penoyre2022b, penoyre2022a}. 
Previous studies of unresolved binaries have generally relied on spectroscopic methods that are most effective for systems exhibiting radial velocity variations, such as identifying double-lined spectroscopic binaries (SB2s) through cross-correlation techniques \citep[e.g.,][]{li2021a} or disentangling component spectra from spectra taken at different orbital phases \citep[e.g.,][]{Seeburger2024}.
These methods are sensitive only to systems with notable velocity separations ($v_{\rm los} \gtrsim 10-30$ km s$^{-1}$) and high mass ratios ($q \ge 0.7$) \citep{zhang2022a}. 
This approach inherently limits detection to relatively short-period binaries, which represent only about one-third of the total binary population \citep{duchene2013}. 

Similarly, the identification of single-lined spectroscopic binaries (SB1s) through radial velocity variability is also biased toward shorter period systems.
Traditional methods are effective within specific period ranges but miss many systems outside their sensitivity windows. 
Spectroscopic techniques, such as identifying single-lined (SB1) and double-lined (SB2) spectroscopic binaries, are limited to systems with periods of approximately 1 to 1000 days, where radial velocity (RV) variations are detectable \citep{el-badry2024}. 
Astrometric methods, on the other hand, are most sensitive to binaries with periods comparable to \gaia’s observational baseline \citep{Castro-Ginard2024}. 
In contrast, Spectral Energy Distribution (SED)-based techniques remain effective across the full range of periods for unresolved binaries with distinct components, as they do not rely on radial velocity or astrometric variability.

This methodology bridges detection gaps, complementing short-period sensitivity of eclipsing binary searches (requiring specific orbital alignments) and long-period capabilities of direct imaging (typically exceeding 10,000 days for Gaia). 
SED-based approaches can detect the estimated 73\% of binaries with periods beyond 10 years \citep{el-badry2018, duchene2013} that remain invisible to conventional RV surveys \citep{gao2014, badenes2018}, while simultaneously providing independent verification for systems within the detection windows of other methods. 
For binaries with periods too long for dynamical detection yet too tight for spatial resolution by \gaia, our XP-based methodology may represent a key identification technique,

The SED method identifies binary stars by modeling the composite spectral energy distribution of two stellar components and detecting deviations from single-star models, with enhanced sensitivity when the components exhibit obvious temperature or luminosity contrasts \citep[e.g.,][]{Cuadrado2001}.
The SED-based methods complement these approaches by operating independently of orbital dynamics, identifying binaries through their composite spectral signatures regardless of period. 
These signals are most pronounced when the two components have different spectral features, as is the case with white dwarf--main-sequence binaries \citep{li2025}. 

However, the presence of an unresolved companion changes the observable spectrum in all cases where the two stars are not identical.
When two main-sequence stars have similar masses, the normalized spectral features or normalized SED show minimal differences between single and binary systems, creating a potential degeneracy \citep{el-badry2018}. 
However, with precise distance measurements (e.g., from \gaia\ parallaxes), this degeneracy can be resolved: a binary system’s total luminosity is the sum of its two components, so for stars with similar spectral features, a binary would appear twice as luminous as a single star. When forced to fit such a binary with a single-star model, the excess luminosity is misinterpreted as a hotter (bluer) temperature to compensate for the unphysically high flux at the known distance.

Together, these complementary detection techniques enable stellar surveys to deliver samples of binary stars and stellar companions orders of magnitude larger than previously known, spanning all stages of stellar evolution.
Even without detectable velocity offsets, unresolved binary spectra contain distinctive signatures of unseen companions. 
By combining the quality of \gaia\, DR3 low-resolution BP/RP (XP) spectra \citep{gaiacollaboration2023} with \gaia's precise parallax measurements \citep{Gaia2016}, we can now independently identify binary systems through chi-squared spectral fitting \citep{li2025}.

The characterization of binary population properties will emerge from integrated analysis across multiple stellar surveys, which collectively sample diverse stellar types, wavelengths, and observational techniques.
This approach addresses gaps in binary detection, complementing both eclipsing binary searches at short periods and direct imaging at long periods, while capturing the estimated 70\% of binaries with periods exceeding 10 years that traditional RV surveys miss \citep{duchene2013,el-badry2018} for binaries with periods too long for dynamical detection yet too close for spatial resolution.

In this paper, we focus on main-sequence main-sequence binary  (MSMS) stars from \gaia\ XP spectra, presenting an inventory of unresolved binary systems in which both components are main-sequence stars.
We employed the high quality of Gaia DR3 low-resolution XP spectra \citep{gaiacollaboration2023}, combined with precise parallax measurements from \gaia, \citep{Gaia2016}, we conduct a homogeneous search for binary systems throughout the Galaxy, identifying and characterizing these systems with completeness and consistency.

Our approach is based on the methodology introduced by \citet{li2025}.
We utilized \gaia\ XP spectra, which provide low-resolution spectrophotometry for over 220 million sources in the Galaxy.
Although these spectra have a lower resolution ($R \sim 20-100$) than traditional spectroscopic surveys, they offer sufficient signal-to-noise and self-consistent flux calibration \citep{deangeli2023}.
By modeling the composite spectral energy distributions of binary components, we can identify binaries regardless of their velocity separation, including long-period systems with negligible velocity offsets between components, provided they have favorable mass ratios ($0.5 \lesssim q \lesssim 1.0$).

In this work, we present a MSMS catalog derived from \gaia\ XP spectra, containing over 14 million binary candidates.
The remainder of the paper is structured as follows: In Section \ref{sec:data}, we describe the \gaia\ DR3 data and our sample selection criteria.
Our methodology, described in Section \ref{sec:method}, builds upon our forward model for stellar spectra mapping intrinsic parameters: stellar mass ($M$), photometric metallicity (\photmoh), and extinction ($E$) to predicted spectra.
In section \ref{sec:validation}, we present the validation results and the method to classify the single and binary stars using parameter-dependent thresholds.
Section \ref{sec:binary_catalog} details the statistical properties of our sample, including our binary candidate selection process and the creation of a high-confidence "golden sample" of approximately 1 million binary systems.
We discuss the implications of our findings and summarize our conclusions in Section \ref{sec:discussion}.
This catalog represents a valuable resource for investigating the formation and evolution of binary systems throughout the Galaxy.
The size and diversity of our sample enables robust statistical analyses of binary system properties across different Galactic environments, offering new insights into the processes governing binary star formation and evolution.

\section{Data}\label{sec:data}
In this section, we first describe the properties and processing of the \gaia\ BP/RP spectra, followed by a detailed outline of our sample selection criteria and pre-processing steps.

\subsection{XP: The Gaia BP/RP Low-Resolution Spectra}\label{subsec:xp}

\gaia, DR3 provides low-resolution spectrophotometric data (XP spectra) for about 220 million sources, obtained with the Blue Photometer (BP, 330--680\,nm) and Red Photometer (RP, 640--1050\,nm) instruments onboard the spacecraft \citep{gaiacollaboration2023}. 
These spectra cover the optical to near-infrared range with a wavelength-dependent resolving power of R$\sim$20-100.

Unlike conventional spectroscopic data consisting of discrete flux measurements, \textit{Gaia} XP spectra implement a continuous functional representation through basis function decomposition \citep{deangeli2023}.  
Each spectrum is parameterized via 55 coefficients derived from orthonormal Gauss-Hermite functions, optimized to capture both broadband morphology and higher-frequency spectral features with minimal information loss.  
For analytical tractability, we transform these coefficient-based representations to wavelength-flux space using {\tt GaiaXPy}\footnote{\href{http://doi.org/10.5281/zenodo.6674521}{http://doi.org/10.5281/zenodo.6674521}}, facilitating direct identification and measurement of diagnostic spectral features as described in Appendix~\ref{app:sample_xp}.  
This transformation yields calibrated flux values in physical units (W\,nm$^{-1}$\,m$^{-2}$) spanning the operational wavelength range (390--1000\,nm).  
The low-resolution XP spectrophotometry provides sufficient spectral discrimination capability for identifying key stellar atmospheric diagnostics \citep{zhang2023a,li2024} while maintaining the statistical power of \textit{Gaia}'s extensive survey volume, rendering it suited for our binary census and characterization program.

\subsection{Sample Selection Criteria}
To construct our parent sample, we execute the following query on the \gaia\ DR3 database:
\begin{lstlisting}
SELECT *
FROM gaiadr3.gaia_source
WHERE parallax > 1
  AND phot_bp_mean_mag < 20
  AND bp_rp BETWEEN 0 AND 5
  AND parallax_over_error > 10
  AND phot_g_mean_mag + 5*log10(abs(parallax)/100) > 3
  AND has_xp_continuous = `true'
\end{lstlisting}
Our sample selection procedure implements multi-stage filtering protocols:
First, we establish a volume-limited sample within a 1\,kpc heliocentric radius (parallax $\varpi > 1$\,mas) containing objects with available BP/RP (XP) continuous spectra and B-R color indices spanning $0 < \mathrm{B-R} < 5.0$, encompassing F-type to M-type stars.
Second, we mitigate systematic flux calibration biases that affect faint red sources in the \textit{Gaia} photometric system by implementing a magnitude threshold of $B < 20$\,mag, where these systematic effects remain below our precision requirements. 
Third, we excise evolved stellar populations by imposing an absolute magnitude criterion $M_G > 3.0$\,mag (un-dereddened), removing giants and supergiants while retaining the main sequence population.
We focus on main-sequence stars to ensure homogeneity in our sample, as their spectral properties are primarily determined by mass and metallicity, simplifying the forward modeling in this study. 
In future work, we plan to incorporate stellar age explicitly, particularly for brighter stars, by forward-modeling spectra with mass, age, and [M/H] as parameters.
Fourth, we enforce an astrometric quality constraints, requiring parallax signal-to-noise ratios $\varpi/\sigma_\varpi > 10$ to ensure accurate absolute magnitude determinations.
The resulting parent sample comprises 35,831,031 unique sources with \gaia\ DR3 identifiers.

\subsection{Training Data}

To build a clean training sample of main sequence stars for our neural emulator of XP spectra, we have implemented a multi-stage filtering process targeting the main sequence within $\sim$333 parsecs (\texttt{parallax}$>3$). 
Our selection methodology prioritizes purity while maintaining a representative population across the lower main sequence.

\begin{figure}
    \centering
    \includegraphics[width=1\linewidth]{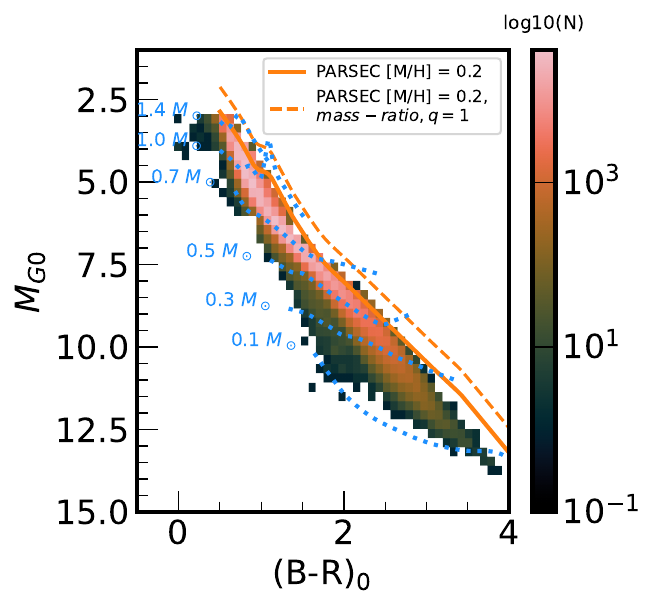}
    \caption{Hertzsprung–Russell diagram of training sample stars. The color scale indicates the logarithm of the number of stars in each bin. 
    The solid orange line represents a PARSEC isochrone with metallicity ([M/H] = 0.2), while the dashed orange line shows the same isochrone for equal-mass binary systems (mass ratio $q = 1$). The blue dotted lines indicate iso-mass tracks, with masses ranging from 0.1 $M_\odot$ to 1.4 $M_\odot$ as labeled.
    }
    \label{fig:hrd_input}
\end{figure}

First, we applied a color-magnitude diagram (CMD) filter using PARSEC theoretical isochrones with parameters of 1 Gyr age and [M/H]$=0.2$ metallicity, and a hard cut of $M_G > 3$ as shown in Figure \ref{fig:hrd_input}. 
By excluding objects above this reference isochrone, we removed most of the main sequence binaries and pre-main sequence objects that would contaminate our sample.
Second, we remove the white dwarf main sequence binary (WDMS) candidates selected from \cite{li2025}.
As a final quality assurance measure, we required all spectra to have an average signal-to-noise ratio greater than 70. 
The resulting clean sample contains 746,548 main sequence stars with high quality XP spectra.

\section{Data-Driven Spectral Model}\label{sec:method}

In this section, we present our data-driven approach for modeling the XP spectra of both single and MSMS stars using neural networks (NN).  
Our methodology encompasses two interconnected components: (1) a single-star spectral emulator that maps stellar parameters directly to observed spectra in Subsection~\ref{subsec:spectral_model}, and (2) a binary star model that combines individual stellar spectra based on fundamental stellar properties in Subsection \ref{subsec:binary_model}. 
Our NN architecture employs a simple yet streamlined design for both single and binary star modeling. 
For single stars, we map photometric stellar parameters directly to spectral flux using a neural network. 
For binary systems, we transform physical parameters into component-specific stellar properties,  and combine the resulting spectra through linear superposition. 

To enhance the capability of modeling single-star XP spectra, we iteratively cleaned the training set three times, removing outliers based on fit quality metrics. 
This process involved evaluating the $\chi^2$ statistic for each spectra in the training set, where high $\chi^2$ values indicated poor agreement between observed XP spectra and the NN's predicted spectra. 
Outliers, defined as spectra with $\chi^2$ values exceeding 3, were removed to ensure the training data consisted of representative single-star spectra. 
Three iterations of this cleaning were performed to refine the dataset progressively, each time retraining the NN to improve its accuracy in mapping stellar parameters.


\begin{figure*}
    \centering
    \includegraphics[width=0.49\linewidth]{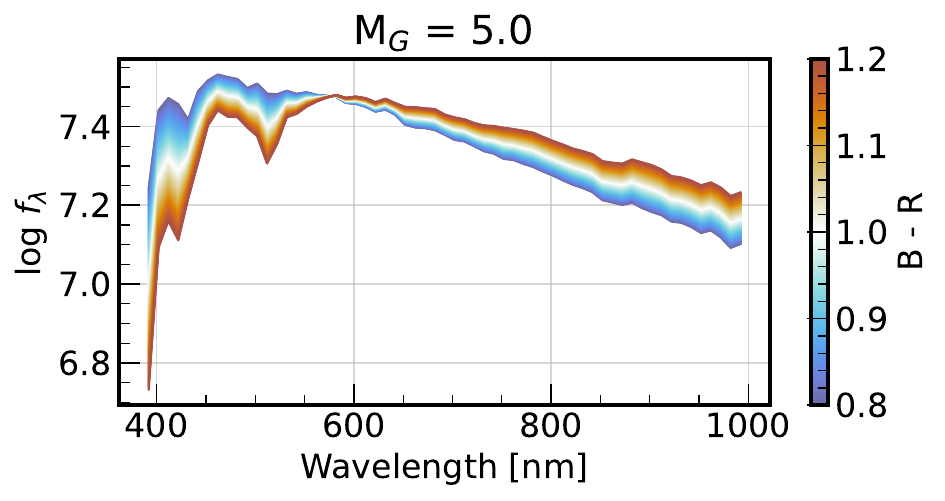}
 \includegraphics[width=0.49\linewidth]{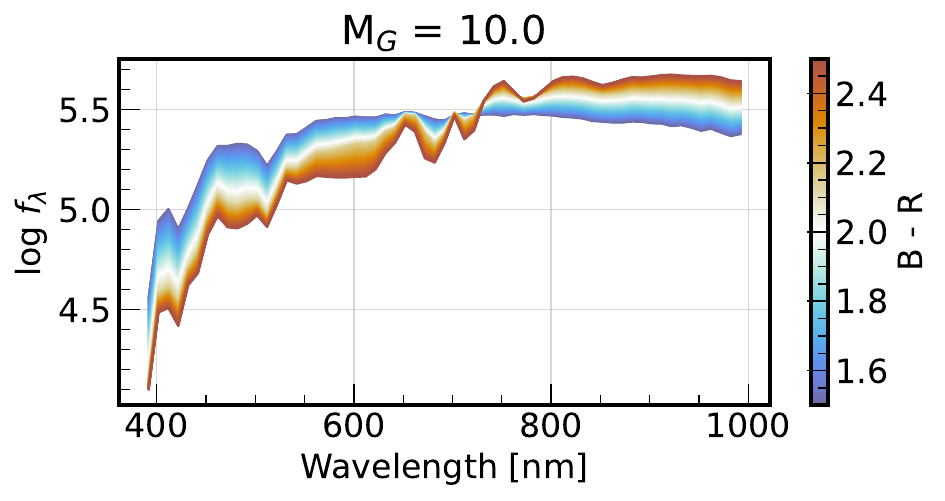}
  \includegraphics[width=0.49\linewidth]{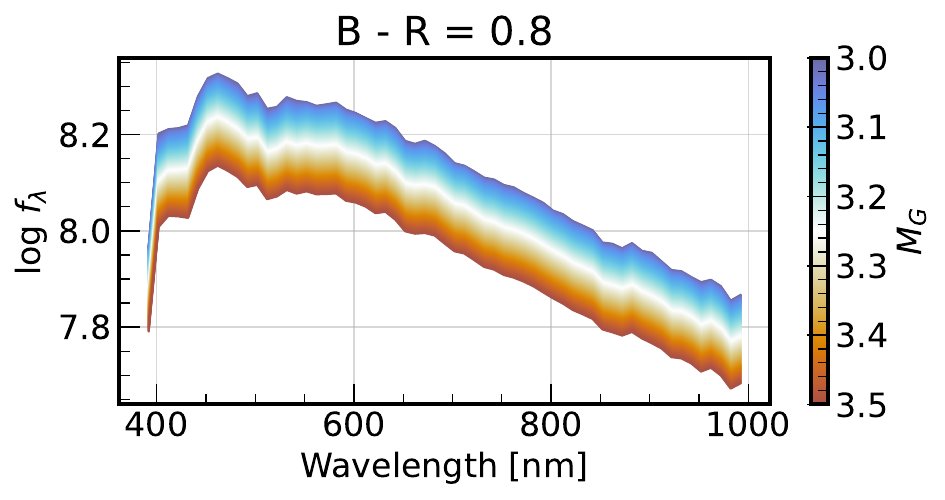}
   \includegraphics[width=0.49\linewidth]{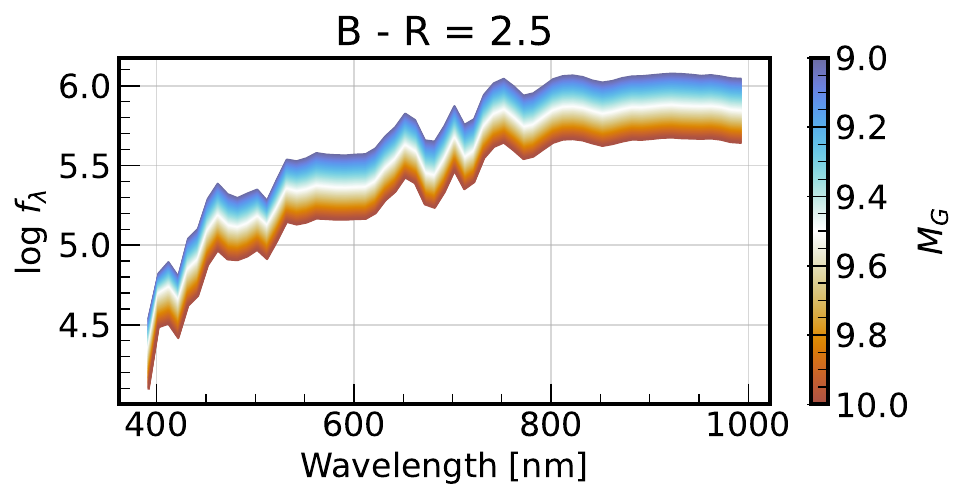}
    \caption{Demonstration of our NN spectral emulators for single MS stars. 
    Each panel shows the predicted XP spectra with different stellar parameters, where the x-axis is the wavelength and the y-axis is the logarithmic flux ($10^{-20}~{\rm W}~{\rm m}^{-2}~{\rm nm}$) normalised to a distance of 10 pc. 
    Top panels: Spectra at fixed absolute G magnitude ($M_G$) with different B-R color. 
    Bottom panels: Spectra at fixed B-R color with varying $M_G$. }
    \label{fig:cmd2xp}
\end{figure*}

We adopt photometric quantities directly related to the Hertzsprung-Russell (H-R) diagram as our stellar labels. 
Specifically, we use
\begin{equation}
    \boldsymbol{\theta} = (B - R, M_G), \end{equation} 
where $B - R$ is the dust-corrected color index, and $M_G$ is the absolute magnitude in the Gaia $G$ band, corrected for extinction using 3D dust maps \citep{Green2018,Edenhofer2024}. 

The de-reddening process follows the method of \cite{li2025}. 
We utilize the \texttt{dustmaps} package \citep{Green2018} to query a 3D dust map and derive the extinction parameter $E$.\footnote{The dust map provides $E$ in units of mag per parsec, which can be converted to extinction at any wavelength by applying the appropriate extinction curve.} 
We adopt the average extinction curve from \cite{zhang2023a}, combined with the transmission curves of the \gaia\ passbands, to compute the extinction corrections.

This choice of $M_G$ is motivated by the fact that $M_G$ provides a reliable and physically meaningful representation of low-mass stars ($M_G \geq 4$--$5$). 
In contrast, $\log g$ is not an ideal descriptor for low-mass stars because it becomes less sensitive to variations in luminosity in this regime. 
In addition, our sample spans a wide range of spectral types, from M-type to G-type stars, making it difficult to define a homogeneous set of $T_{\text{eff}}$ labels. 
The parameters ($T_{\text{eff}}$, $\log g$, [M/H]) of M-type stars are often parameterized differently from previous spectral types (e.g., \citealt{jdli2021}; \citealt{Qiu2024}), leading to inconsistencies when using $T_{\text{eff}}$ as a universal label. 
Furthermore, our approach is purely data-driven, relying on observational data rather than theoretical assumptions. 
By focusing on photometric quantities such as $B - R$ and $M_G$, we ensure that our model is both reliable and applicable to the full range of stellar types in our sample.

\subsection{Spectral Emulator}\label{subsec:spectral_model}

The neural network architecture is designed to be as simple as possible, consisting of three hidden layers, each containing 16 neurons activated by Rectified Linear Units (ReLU). 
The configuration optimizes computational efficiency and spectral modeling.
While XP spectra in logarithmic absolute flux space exhibit approximately linear relationships with absolute magnitude and colors, the spectral features and detailed line profiles require nonlinear transformations to accurately represent the underlying physics of stellar atmospheres. 

The neural network defines a function $f$ that maps from stellar parameters to the predicted XP spectrum:
\begin{equation}
    f(\lambda; \boldsymbol{\theta}, \boldsymbol{w}) = \mathbf{F}(\lambda),
\end{equation}
where $\mathbf{F}(\lambda)$ represents the flux at wavelength $\lambda$, $\boldsymbol{\theta}$ are the stellar parameters, and $\boldsymbol{w}$ denotes the trainable parameters of the network. 

We train the network using the Adam optimizer with a learning rate of $10^{-4}$ and a batch size of 16,384. 
The loss function is defined as the chi-squared ($\chi^2$) statistic, which measures the discrepancy between the observed and predicted spectra:
\begin{equation}
\begin{split}
    \mathcal{L} &= \frac{1}{N} \sum_{i=1}^N \chi_i^2 \\
    &= \frac{1}{N} \sum_{i=1}^N (\mathbf{F}_i^{\text{obs}}(\lambda) - f(\lambda; \boldsymbol{\theta}_i, \boldsymbol{w}))^T \mathbf{C}_i^{-1} (\mathbf{F}_i^{\text{obs}}(\lambda) - f(\lambda; \boldsymbol{\theta}_i, \boldsymbol{w})),
\end{split}
\end{equation}
where $\mathbf{F}_i^{\text{obs}}(\lambda)$ represents the observed spectrum of the $i$-th star, $\mathbf{C}_i$ is the covariance matrix for the $i$-th observation, and $N$ is the total number of stars in the training set.

\subsection{Binary Spectral Model}
\label{subsec:binary_model}

\begin{figure*}
    \centering
    \includegraphics[width=0.49\linewidth]{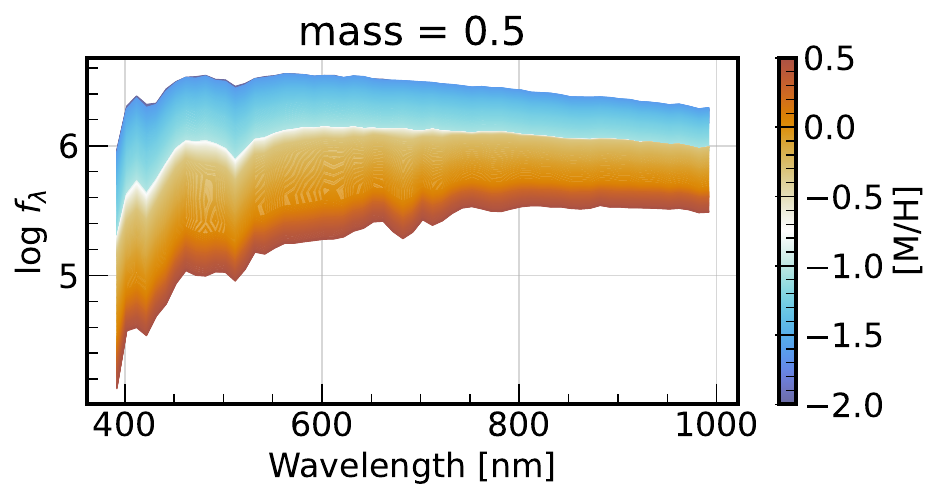}
    \includegraphics[width=0.49\linewidth]{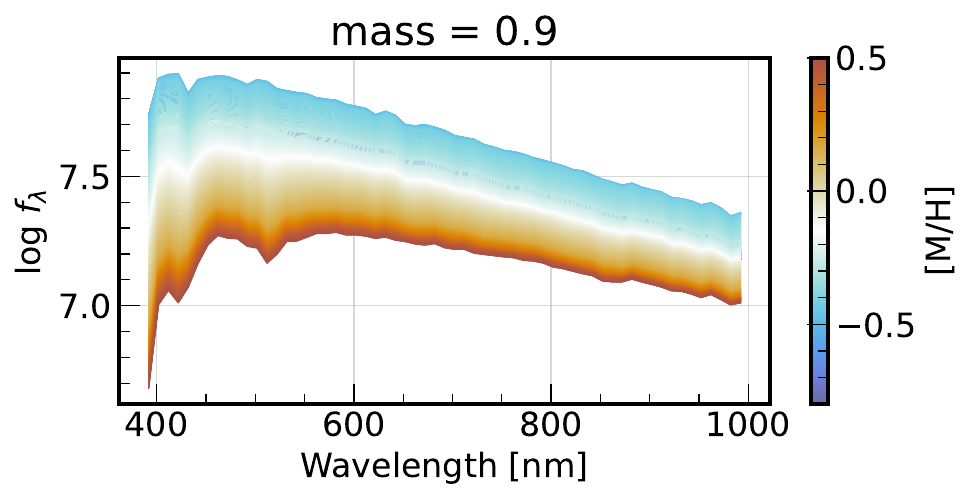}
    \includegraphics[width=0.49\linewidth]{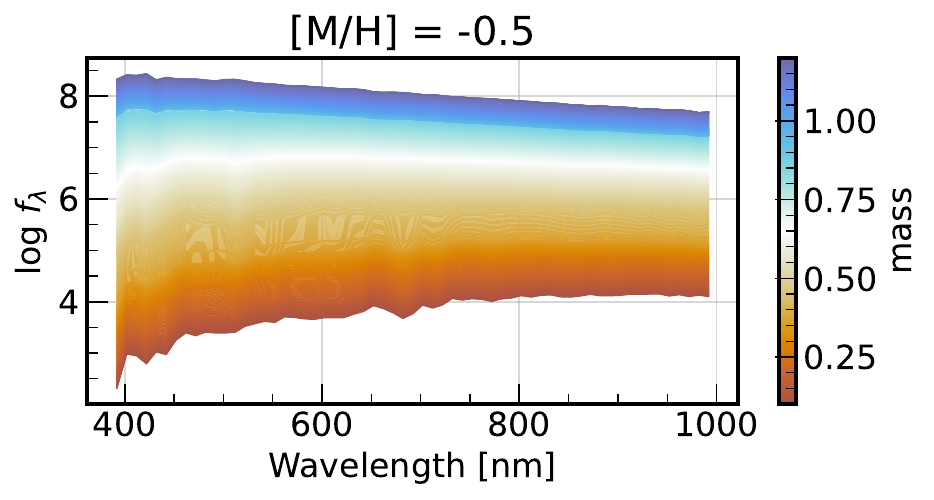}
    \includegraphics[width=0.49\linewidth]{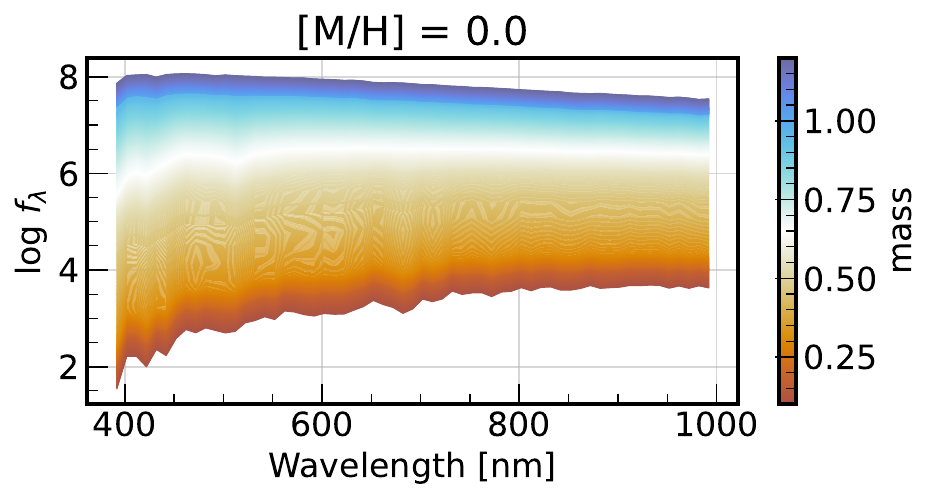}
    \caption{Demonstration of NN spectral emulators for single MS stars. 
    Top panels: Spectra at fixed stellar mass with different \photmoh. 
    Bottom panels: Spectra at fixed \photmoh with varying stellar mass.}
    \label{fig:xp-mass}
\end{figure*}

\subsubsection{Binary System Parameterization}

For modeling binary star systems, we perform stellar parameterization based on stellar properties.

Each binary system is characterized by three primary parameters:
\begin{equation}
    \boldsymbol{\phi} = (m_1, \text{[M/H]}_{\text{photo}}, q),
\end{equation}
where $m_1$ represents the mass of the primary star, \photmoh\ is the metallicity (assumed identical for both components), and $q = m_2/m_1$ is the mass ratio, constrained to $0 < q \leq 1$. 
This parameterization captures the key properties of binary systems while minimizing the number of free parameters. 

\subsubsection{Mapping Stellar Parameters to the H-R Diagram}
To translate the physical parameters into observable quantities on the H-R diagram, we use the PARSEC stellar evolution models \citep{Bressan2012,Chen2014} with stellar ages of 5 Gyr.
We implement this mapping using a neural network function $\mathcal{M}$ that captures the nonlinear transformations:
\begin{equation}
    \boldsymbol{\theta} = \mathcal{M}(m_1, \text{[M/H]}_{\text{photo}}),
\end{equation}
where $\boldsymbol{\theta} = (B - R, M_G)$ represents the position in the H-R diagram expressed as dust-corrected color index and absolute magnitude.

The photometric properties of each component in a binary system are computed separately:
\begin{align}
    \boldsymbol{\theta}_1 &= \mathcal{M}(m_1, \text{[M/H]}_{\text{photo}}), \\
    \boldsymbol{\theta}_2 &= \mathcal{M}(q, m_1, \text{[M/H]}_{\text{photo}}).
\end{align}
Here, $\text{[M/H]}_{\text{photo}}$ denotes the photometric metallicity derived from broadband colors. 
The photometric metallicity reflects the integrated effect of metal content on the stellar atmosphere's opacity and emergent flux, which may differ from element-specific spectroscopic determinations.
Full details of this mapping implementation and its validation are provided in Appendix~\ref{app:hr_mapping}. 

\subsubsection{Combined Binary Spectrum}
Once the H-R diagram positions of both components are determined, we predict the individual spectra using the same mapping function $f$ from Section~\ref{subsec:spectral_model}:
\begin{align}
    \mathbf{F}_1(\lambda) &= f(\boldsymbol{\theta}_1), \\
    \mathbf{F}_2(\lambda) &= f(\boldsymbol{\theta}_2).
\end{align}

The combined spectrum of the unresolved binary system is then computed as the sum of the individual component spectra:
\begin{equation}
    \mathbf{F}_{\text{binary}}(\lambda; m_1, \text{[M/H]}_{\text{photo}}, q) = \mathbf{F}_1(\lambda) + \mathbf{F}_2(\lambda).
\end{equation}

This additive model assumes the flux contributions from the two stars are linearly superimposed, which is a valid approximation for non-interacting binary systems.

\begin{figure*}
    \centering
    \includegraphics[width=0.49\linewidth]{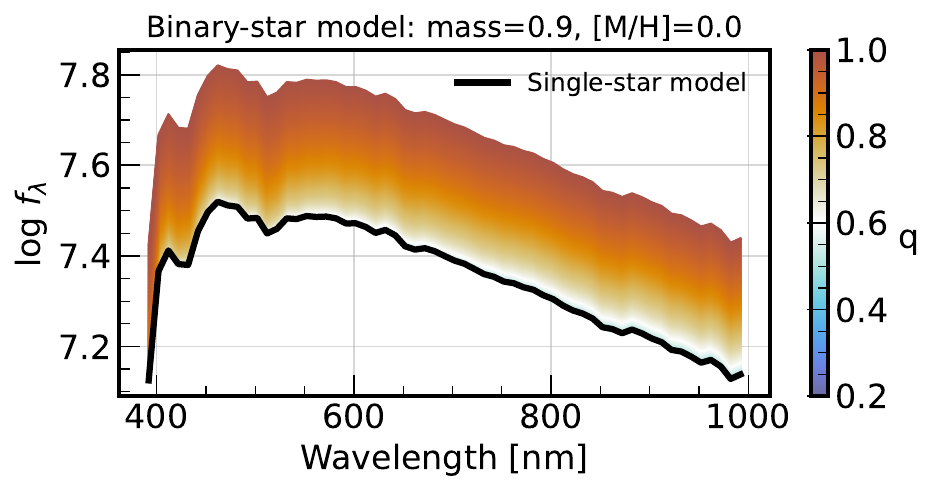}
    \includegraphics[width=0.49\linewidth]{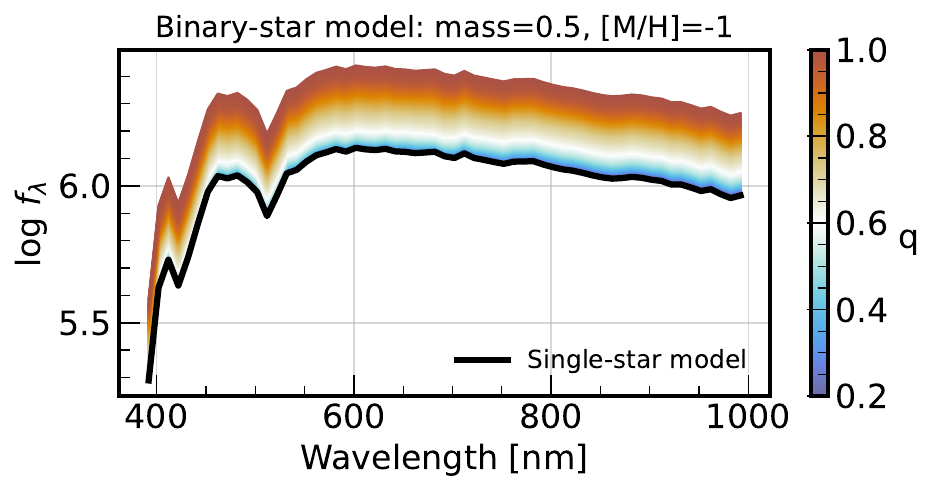}
    \label{fig:xp-mass-binary}
    \caption{Comparison of single-star and binary-star XP spectral models for systems with varying mass ratios. 
    Black lines represent the single-star model, while the color gradient represents binary-star models with varying mass ratios ($q = m_2/m_1$), ranging from $q=0.2$ (blue) to $q=1.0$ (red). The binary models incorporate consistent \photmoh, and distance for both components.}
\end{figure*}

\subsubsection{Extinction Modeling}
\label{subsubsec:extinction_model}
We model extinction as a wavelength-dependent attenuation of stellar flux following standard extinction laws. 
The extinction is characterized by a parameter $E$, which scales the universal extinction curve $R(\lambda)$. 
The observed flux is related to the intrinsic flux by:
\begin{equation}
    \mathbf{F}_{\text{obs}}(\lambda) = \mathbf{F}_{\text{intrinsic}}(\lambda) \cdot \exp(-E \cdot R(\lambda)),
\end{equation}
where $R(\lambda)$ represents the relative extinction at each wavelength. 
For our analysis, we adopt the average extinction curve derived by \cite{zhang2023a}, which was determined from a large sample of \textit{Gaia} XP spectra spanning diverse Galactic environments.

\subsubsection{Parallax Error Propagation}
\label{subsec:parallax_error_prop}

We normalize all XP spectra to the flux at 10 pc using the parallax from Gaia DR3. 
Since the parallax measurement has uncertainties that cannot be neglected, we perform error propagation as follows:  
\begin{equation}  
\vec{f}_{10} = \vec{f}_{\mathrm{obs}} \left( \frac{100}{\varpi} \right)^2  
\end{equation}  
where \(\vec{f}_{10}\) is the flux normalized to 10 pc, \(\vec{f}_{\mathrm{obs}}\) is the observed flux at the star's distance, and \(\varpi\) is the parallax in milliarcseconds (mas).  
The error propagation process accounts for the uncertainty in parallax measurements when normalizing XP flux spectra to a standard distance of 10 pc.

We propagate the errors using two components:
\begin{equation}
C_{\vec{f}_{10}} = g_\varpi^2 C_{\vec{f}_{\rm obs}} + J J^T \sigma_\varpi^2,
\end{equation}
where $g_\varpi$ is defined as the scaling factor:
\begin{equation}
g_\varpi = \left(\frac{100}{\varpi}\right)^2;
\end{equation}
The first term scales the original covariance matrix ($C_{\vec{f}_{\rm obs}}$) by the square of the scaling factor, while the second variance term accounts for parallax uncertainty. 
This second term is computed using the Jacobian ($J$) of the scaling function with respect to parallax, multiplied by the square of the parallax error ($\sigma_\varpi$):
\begin{equation}  
J = \frac{\partial g_\varpi}{\partial \varpi} \vec{f}_{\mathrm{obs}} = -\frac{20000}{\varpi^3} \vec{f}_{\mathrm{obs}}  
\end{equation}  

The resulting covariance matrix ($C_{\vec{f}_{10}}$) incorporates both the scaled original uncertainties and the additional uncertainty introduced by the parallax measurement. 
This error propagation ensures that measurement uncertainties are properly accounted for in our model fitting and parameter inference.

\subsection{Maximum Likelihood Parameter Estimation}
\label{subsec:parameter_estimation}

We employ a likelihood-based parameter inference of both single and binary stellar systems with a framework below.

\subsubsection{Likelihood Formulation}
Assuming Gaussian uncertainties in the observed fluxes, we define the likelihood function as:
\begin{equation}
    \mathcal{L}(\boldsymbol{\Phi}) \propto \exp\left(-\frac{1}{2} \chi^2(\boldsymbol{\Phi})\right),
\end{equation}
where $\boldsymbol{\Phi}$ represents either single-star parameters $(\boldsymbol{\theta}, E)$ or binary parameters $(m_1, \text{[M/H]}_{\text{photo}}, q, E)$. The $\chi^2$ statistic is defined as:
\begin{equation}
    \chi^2(\boldsymbol{\Phi}) = \Delta\mathbf{F}^T C_f^{-1} \Delta\mathbf{F},
\end{equation}
where $\Delta\mathbf{F}$ is the vector of residuals between observed and model-predicted fluxes, and $C_f$ is the covariance matrix of flux uncertainties.

\subsubsection{Modified Likelihood with Prior Information}
To incorporate prior information about extinction, we maximize a modified likelihood function:
\begin{equation}
    \ln \mathcal{L}_{\text{mod}}(\boldsymbol{\Phi}) = -\frac{1}{2} \chi^2(\boldsymbol{\Phi}) - \frac{1}{2} \lambda_p \cdot (E - E_{\text{ref}})^2,
\end{equation}
where the second term functions as a penalty encouraging the extinction parameter $E$ to remain close to the reference value $E_{\text{ref}}$ from the extinction map \citep{Edenhofer2024}. 
The parameter $\lambda_p$, set to 0.05, controls the strength of the extinction constraint in the modified likelihood function. 
This term acts as a regularization, anchoring the fitted extinction $E$ to the reference value $E_{\text{ref}}$ from the dust map \citep{Edenhofer2024}.

\subsubsection{Optimization Procedure}
For parameter estimation, we employ the Trust Region Reflective (TRF) algorithm \citep{coleman1994centering}, which is effective for constrained optimization problems. 
The TRF method constructs a local quadratic approximation of the objective function within a trust region, ensuring stability and avoiding issues such as overstepping or divergence that can occur with other optimization methods.

The parameter space exploration is bounded by physical constraints:
\begin{itemize}
    \item Primary mass $m_1 \in [0.1,\ 1.5]\ M_\odot$
    \item Photometric metallicity $[\text{M/H}]_\text{phot} \in [-2.5,\ 0.7]$\;dex
    \item Mass ratio $q$ strictly satisfies $0.1 < q < 1.0$
\end{itemize}

The optimal parameters are obtained by solving:
\begin{equation}
    \hat{\boldsymbol{\Phi}} = \arg\min_{\boldsymbol{\Phi}} \left(-\ln \mathcal{L}_\text{mod}(\boldsymbol{\Phi})\right)
\end{equation}
In our implementation, we compute the residual vector and extinction penalty separately, then combine them into an augmented residual vector. 
This allows us to treat the extinction penalty as an additional ``observation'' in our least-squares formulation while rigorously enforcing the parameter boundaries through the TRF algorithm's constraint-handling capabilities.

\begin{figure*}
    \centering
    \includegraphics[width=1\linewidth]{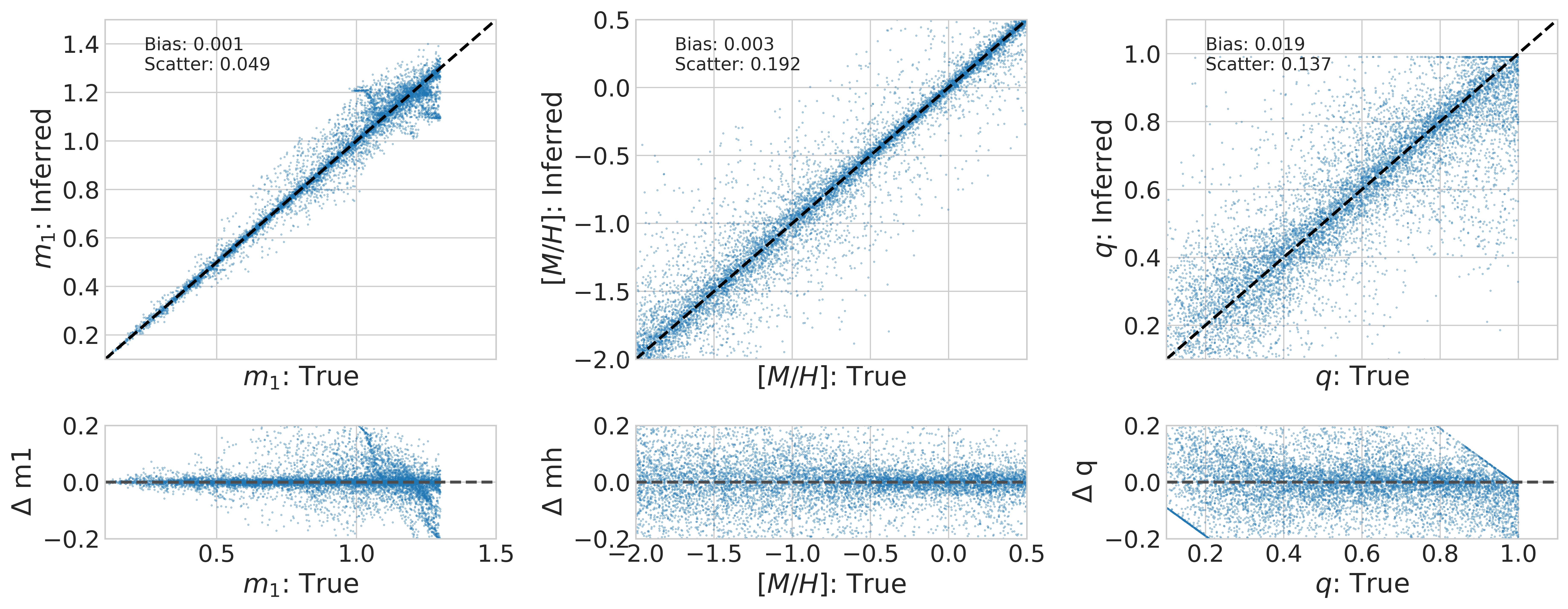}
    \caption{Validation of parameter recovery for the MSMS mock data. Top row: Comparison between true and inferred values for primary mass ($m_1$, left), \moh ([M/H], middle), and mass ratio ($q$, right). 
    The black dashed lines show the one-to-one relation, while blue points represent individual systems in our validation sample. Inset text shows the median bias and scatter for each parameter. Bottom row: Residuals (inferred minus true values) as a function of the true parameter values, highlighting the consistency of our parameter recovery across the full range of stellar properties. 
    Primary masses are recovered with high precision (scatter of 0.049 $M_\odot$), while \moh\ show broader but still reliable recovery (scatter of 0.192 dex). Mass ratios are determined with good precision (scatter of 0.137) with a small positive bias (0.019) at low $q$ values because of the detection sensitivity limitations. 
    The hard limits in the right panel result from enforcing $q < 1$.
    }
    \label{fig:compare_qfit}
\end{figure*}

\begin{figure}
    \centering
    \includegraphics[width=0.8\linewidth]{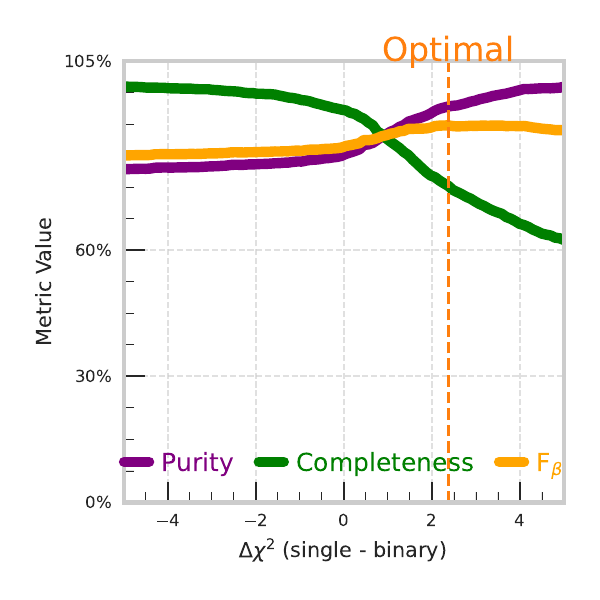}
\caption{Performance metrics as a function of the difference between $\chi^2$ single and $\chi^2$ binary values. 
The plot shows purity (precision, purple line), completeness (recall, green line), and F$_\beta$ score (orange line) across different threshold values, where $\beta =0.5$.
The vertical dashed line is the $\chi^2$ threshold that maximum the $F_{\beta}$ score.
 Beyond this threshold, completeness decreases most rapidly, followed by F$_\beta$ score, while purity remains high. 
 }
    \label{fig:pr_curve}
\end{figure}

\section{Validation}\label{sec:validation}

\begin{figure*}
    \centering
    \includegraphics[width=0.49\linewidth]{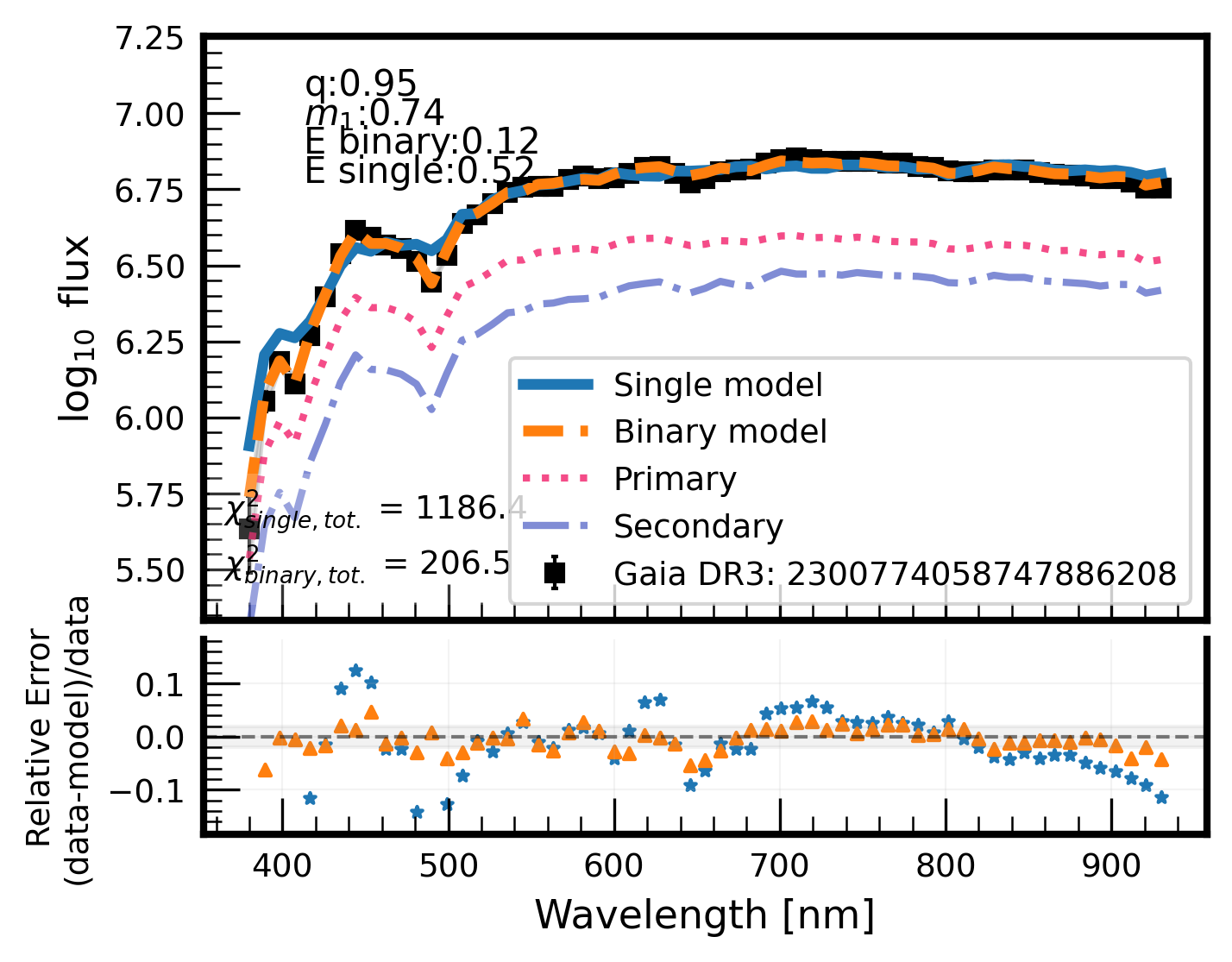}
    \includegraphics[width=0.49\linewidth]{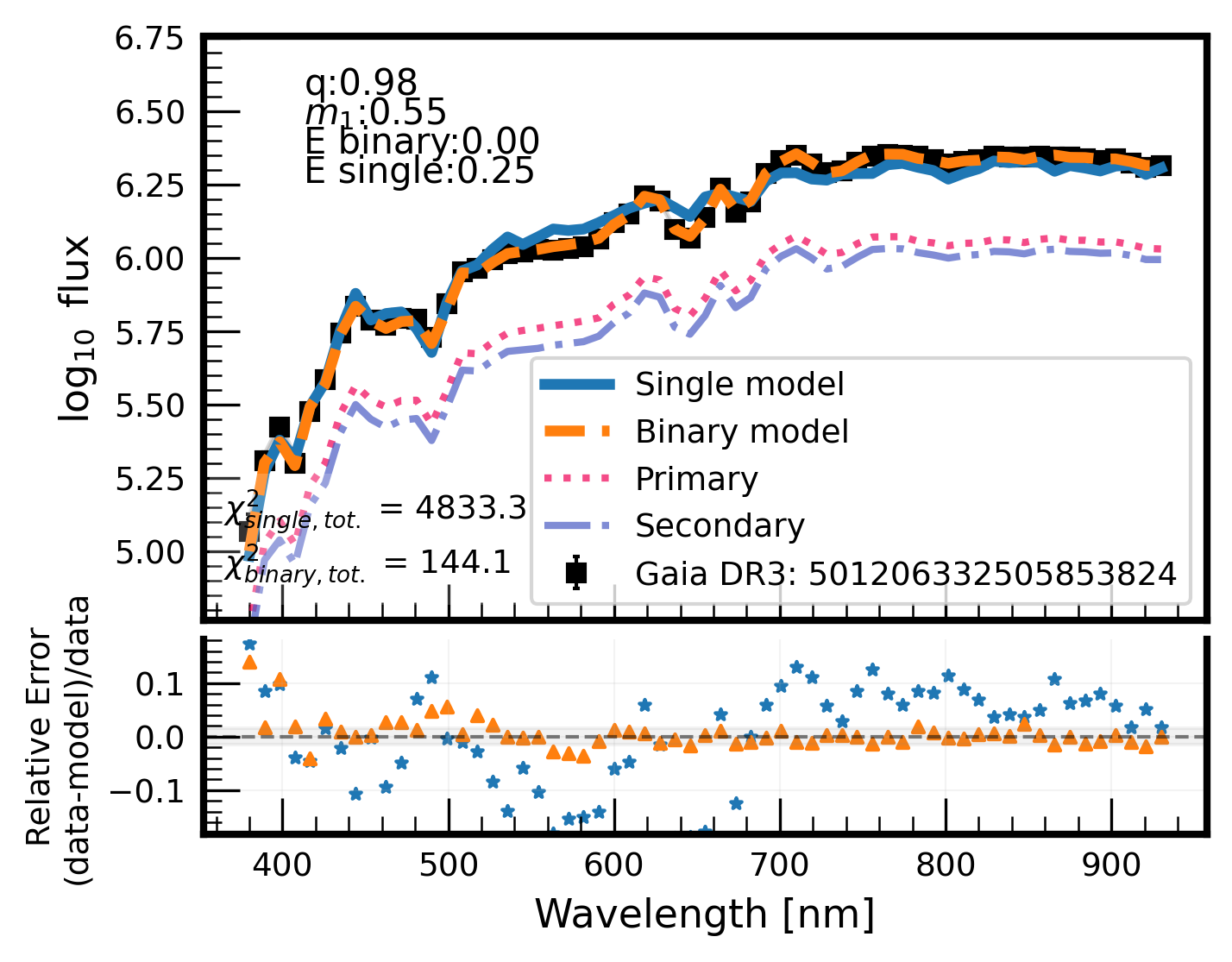}
    \includegraphics[width=0.49\linewidth]{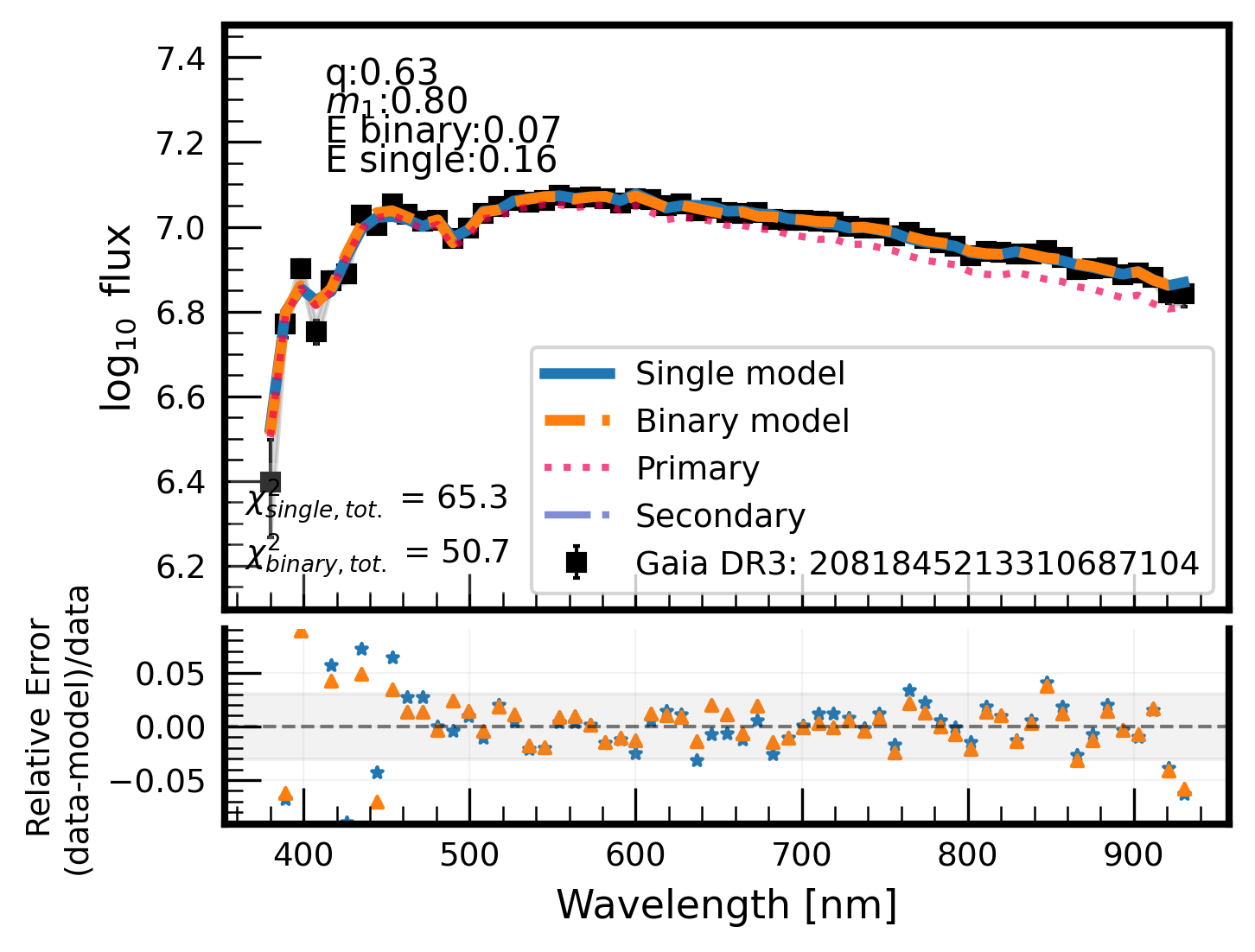}
    \includegraphics[width=0.49\linewidth]{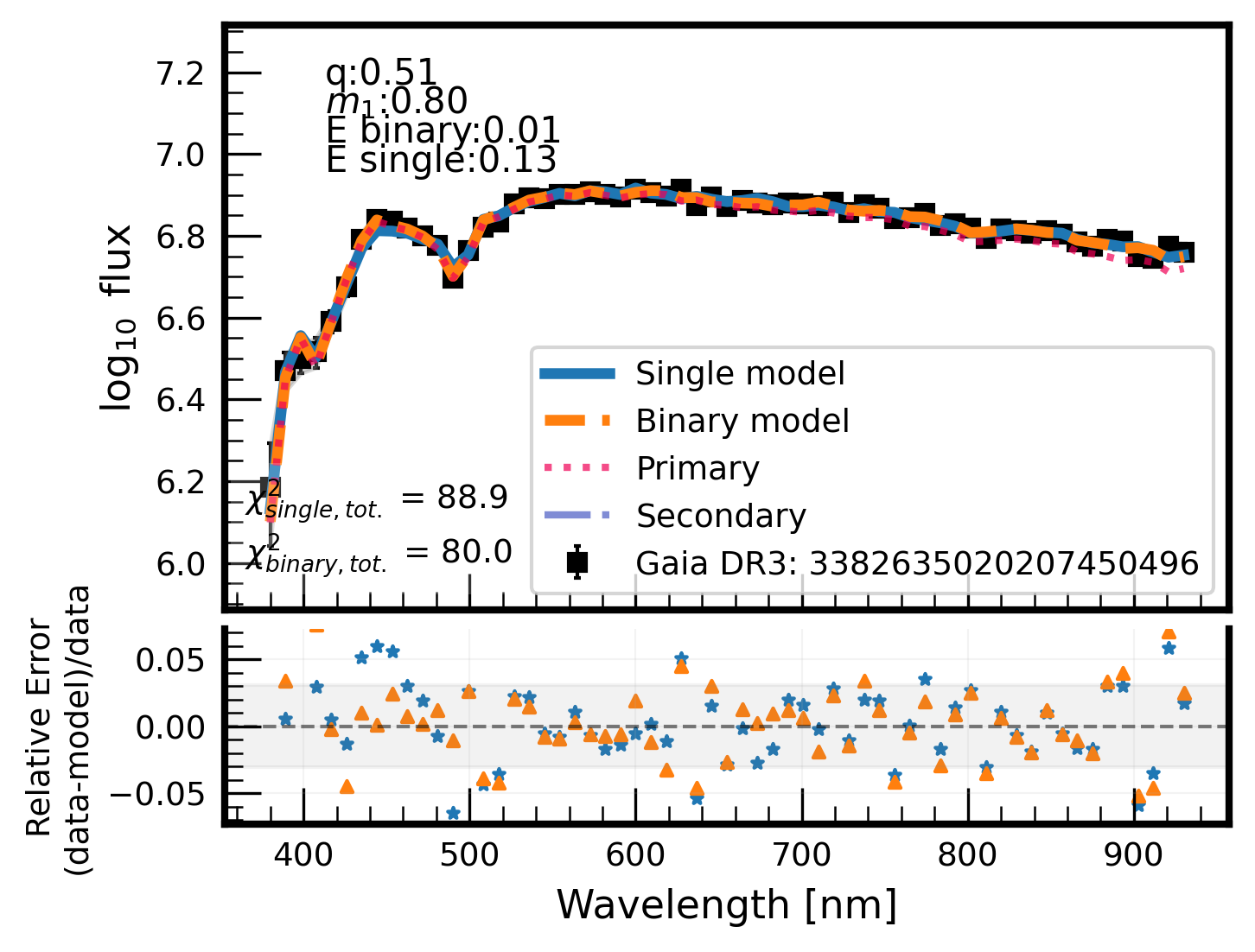}
    \caption{
       \gaia\ XP spectra and best-fitting models for four MSMS binaries.
       In each panel, the black line represents the observed flux.
       The blue solid line shows the best-fit single-star model, while the orange dashed line represents the best-fit binary model. 
       The pink dotted line corresponds to the primary component of the binary model, and the dark blue dotdash line shows the secondary component.
       For each system shown, the binary model is preferred as it achieves a lower $\chi^2$ value than the single-star model. 
       Top row: Example NSS SB2 binary. Bottom row: Example NSS astrometric binaries. 
    }
    \label{fig:sb2_fit}
\end{figure*}

To validate our binary parameter recovery method, we constructed a suite of mock binary systems designed to span the full range of expected stellar parameters. 
Since we already have the single star emulator as described in section \ref{sec:method}, we will first generate stellar labels and then use the single star emulator to mock the single and binary XP spectra.
We generated 10,000 synthetic binary systems using a uniform random distribution for primary masses ($m_1$) ranging from 0.1 to 1.3 $M_\odot$ and \photmoh\ ranging from $-2.0$ to $+0.5$ dex. 
Secondary masses ($m_2$) were similarly drawn from a uniform distribution between 0.1 and 1.3 $M_\odot$.  
We then performed random pairing of primary and secondary stars, followed by reordering when necessary to ensure $m_1 > m_2$ in all cases, resulting in mass ratios ($q = m_2/m_1$) spanning the full physical range from 0.1 to 1.0. 
For each synthetic binary, we generate
mock \gaia\ XP spectra from the XP emulator with Gaussian noise properties based on the Gaia XP error model, incorporating the appropriate signal-to-noise. 
These mock spectra were then processed through our parameter inference pipeline, allowing us to quantify biases and scatter in the recovery of primary mass, metallicity, and mass ratio across the entire parameter space, as shown in Figure~\ref{fig:compare_qfit}.

Figure \ref{fig:compare_qfit} demonstrates the accuracy of our parameter recovery for mock binary systems at S/N=50. 
The top panels show the correlation between true and inferred parameters for primary mass ($m_1$), [M/H]$_{\rm phot}$, and mass ratio ($q$), while the bottom panels display the residuals. 
Primary mass recovery is excellent with minimal bias (0.001) and scatter (0.049), showing that our model accurately constrains the dominant component. [M/H]$_{\rm phot}$ is also well recovered with only slight bias (0.003) and moderate scatter (0.192), which is reasonable given the metallicity's more subtle spectral effects. 
The mass ratio exhibits somewhat larger scatter (0.137) and bias (0.019), particularly for low $q$ values where the secondary's spectral contribution is minimal. 
The residuals show no systematic trends across the parameter ranges, confirming our binary model's robustness for systems with moderate to high signal-to-noise ratios.

 The best-fit masses show larger scatter for stars more massive than 1 $M_\odot$, as these stars evolve on the HR-diagram during their main-sequence lifetime. 
 This highlights that we cannot neglect the effects of stellar ages for more massive stars. 
 In future work, we will further refine our approach by explicitly modeling colors and magnitudes as functions of mass, age, and metallicity to better account for evolutionary effects.

\subsection{Classification Methodology}
To identify binary systems, we compare the goodness-of-fit between single-star and binary models using the $\chi^2$ statistic. A source is classified as a binary if:
\begin{equation}
    \Delta \chi^2 = \chi^2_{\text{single}} - \chi^2_{\text{binary}} > S,
\end{equation}
where $S$ represents the optimal threshold for binary classification.

The threshold $S$ is calibrated to balance completeness and purity, with an emphasis on minimizing false positives. We optimize using the $F_\beta$-score with $\beta = 0.5$:
\begin{equation}
    F_\beta = (1 + \beta^2) \cdot \frac{\text{Purity} \cdot \text{Completeness}}{(\beta^2 \cdot \text{Purity}) + \text{Completeness}}.
\end{equation}

Setting $\beta < 1$ explicitly prioritizes purity over completeness, reducing contamination from single stars incorrectly classified as binaries.

\subsection{Parameter-Dependent Thresholds}\label{subsec:threshold}
We determine optimal thresholds within bins of primary mass ($m_1$) and \photmoh\  
and mass ratio ($q$) rather than applying a uniform threshold across all stellar types. The optimal threshold $S$ varies across these parameters as shown in Table~\ref{tab:chi2_thresholds}, reflecting how binary detectability depends on stellar properties. 

For validation, we generate a mock dataset of 100,000 spectra (50,000 single and 50,000 binary) using our spectral emulators. We randomly sample 10,000 stars with primary masses ranging from 0.1 to 1.5 $M_{\odot}$, photometric metallicities ([M/H]$_{\text{photo}}$) from -2.0 to +0.5 dex, and for binary systems, mass ratios ($q$) from 0.1 to 1.0. 
Each spectra is then simulated at five different signal-to-noise ratios (S/N = 5, 10, 20, 50, 100) to assess classification performance across varying data quality conditions. We add Gaussian noise to the synthetic fluxes according to these S/N levels to realistically mimic observational uncertainties. 

For each mock spectrum, we perform both single-star and binary model fits to obtain best-fit parameters $\boldsymbol{\Phi}_{\text{single}}$ and $\boldsymbol{\Phi}_{\text{binary}}$, then calculate the $\chi^2$ values and their difference ($\Delta\chi^2$). 
Within each bin of primary mass, metallicity, and mass ratio, we determine the threshold $S$ that maximizes the $F_\beta$-score. 
Figure \ref{fig:pr_curve} shows the performance metrics as a function of $S$ for all mock dataset. 
The optimal point is set to maximize the $F_\beta$ value, which in this case is 2.37. 
This approach ensures our classification accounts for the varying sensitivity of the $\Delta\chi^2$ statistic across different regions of parameter space. 

Binaries with nearly equal-mass components typically require lower thresholds for identification as they produce more spectral deviations from single-star models. 
Conversely, systems with low mass ratios require more stringent thresholds to avoid false positives,. 
The resulting parameter-dependent threshold map in Table~\ref{tab:chi2_thresholds} enables binary classification across the entire dataset while balancing purity and completeness.

\subsection{Verification with Binary Catalogs}\label{subsec:verification_gnss}
\begin{figure*}
    \centering
    \includegraphics[width=0.34\linewidth]{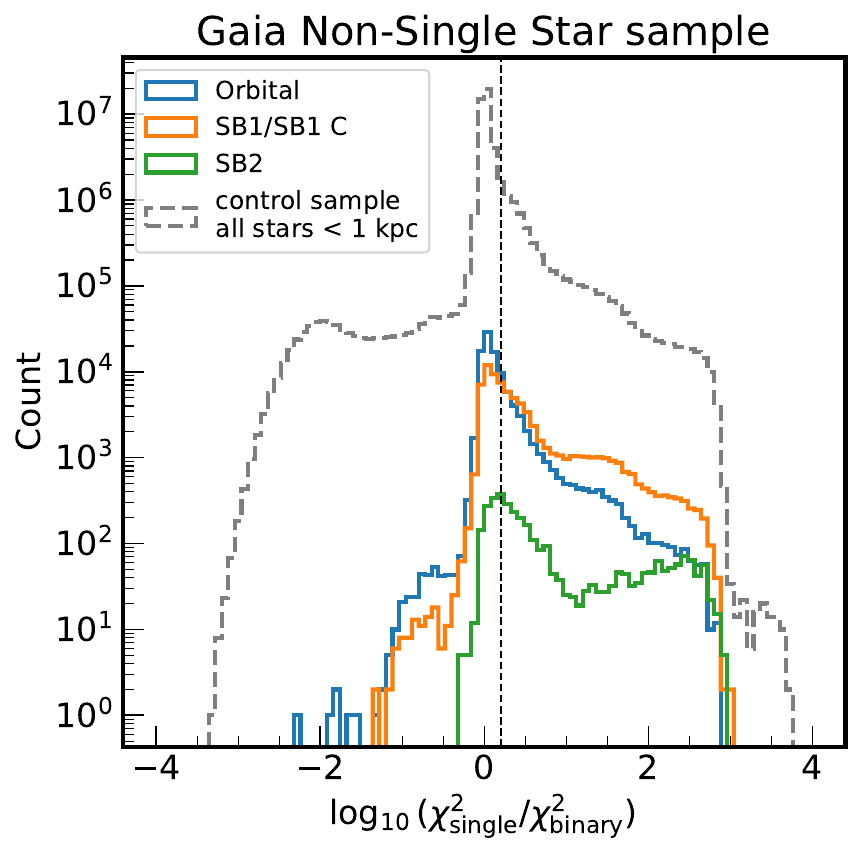}
    \includegraphics[width=0.32\linewidth]{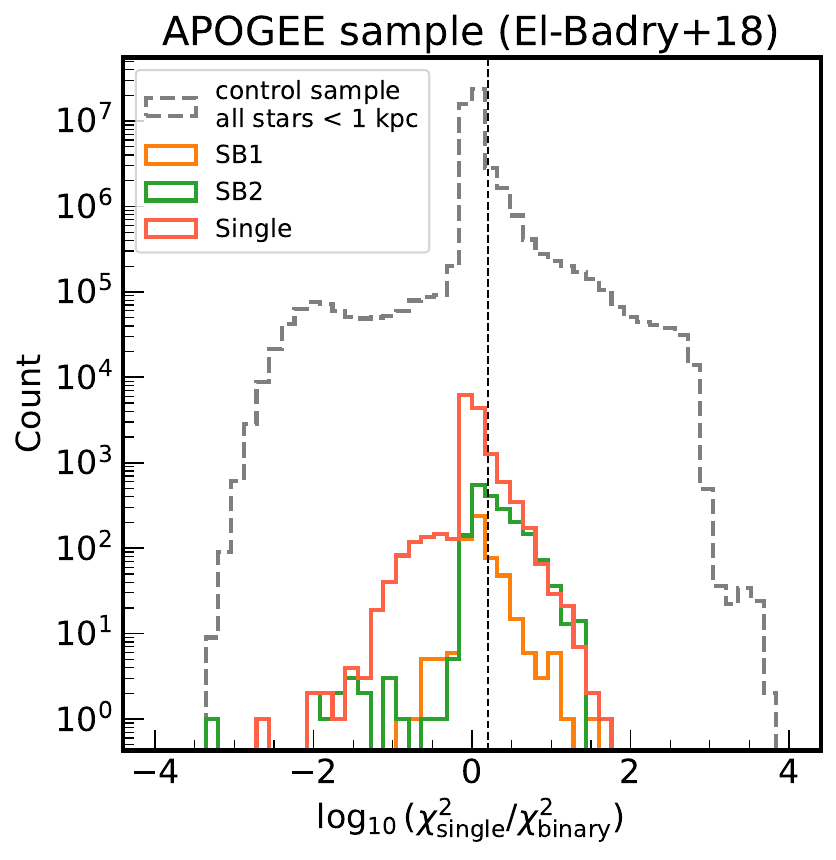}
    \includegraphics[width=0.33\linewidth]{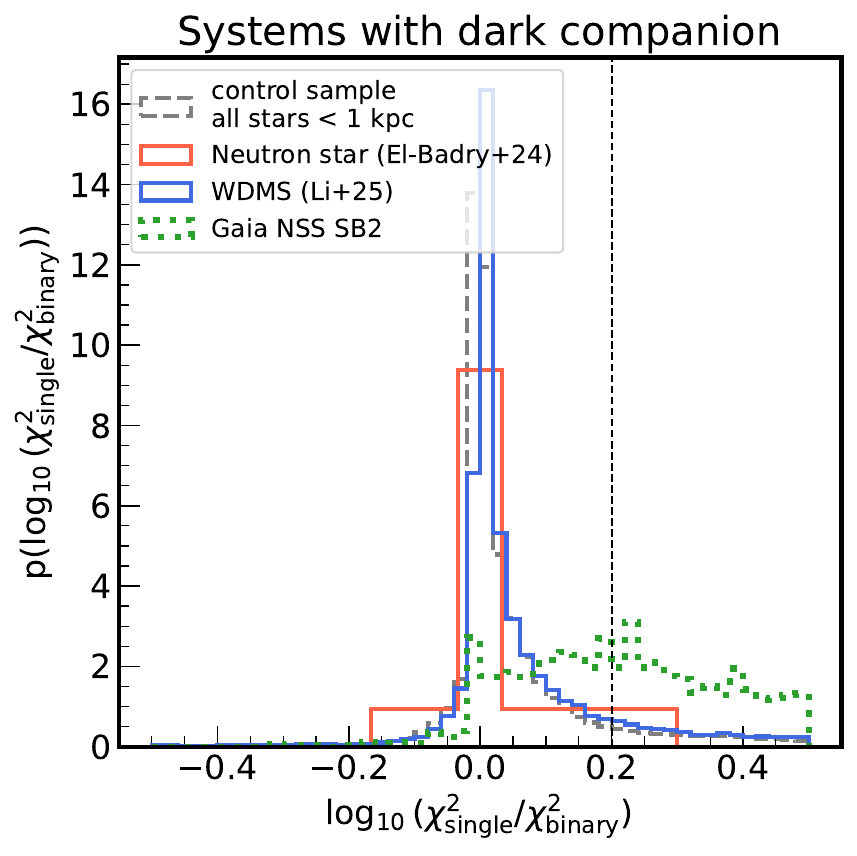}
\caption{Distribution of the logarithmic $\chi^2$ improvement, $\log_{10}(\chi^2_{\text{single}} / \chi^2_{\text{binary}})$, showing the statistical evidence favouring binary over single-star spectral models.
Left panel: Distributions for Gaia DR3 Non-Single Star (NSS) sources: systems with orbital solutions (blue), single-lined spectroscopic binaries (SB1/SB1C; orange), and double-lined spectroscopic binaries (SB2; green).
Right panel: Distributions for specialized populations with likely dark companions: neutron star binaries \citep{el-badry2024a} (red) and white dwarf-main sequence binaries (WDMS) \citep{li2025} (blue). The Gaia DR3 NSS SB2 sample (green dotted line) is shown for reference.
Common elements in each panel: The distribution for all stars within 1 kpc (black dashed line) acts as a control representing the general population. A vertical dashed line marks 0.2 for reference. Higher values of the $\chi^2$ improvement indicate stronger evidence for binarity; known binaries typically show significant positive values.}
\label{fig:chi2_diff_gnss_dark}
\end{figure*}
\begin{table}[ht]
\centering
\caption{Fraction of Systems labeled as binaries by adopted threshold}
\begin{tabular}{l c c cl}
\hline
Dataset & Astrometric/Orbital Binary & SB1 & SB2  &single\\
\hline
Gaia NSS & 67\%& 81\%& 93\%&--\\
APOGEE & -- & 52\%& 85\%&34\%\\
\hline
\end{tabular}
\label{tab:binary_fractions}
\end{table}

The \gaia\ DR3 Non-Single Star (NSS) catalog \citep{gaiacollaboration2023} identified 813,687 multiple systems, including 169,227 astrometric orbital solutions, 220,372 spectroscopic solutions, 87,073 eclipsing binaries, and 338,215 acceleration solutions \citep{gaiacollaboration2023a}.
Spectroscopic binaries are most detectable at periods of 1-1000 days, limited by stellar rotation at short periods and the DR3 baseline at longer periods. 
The NSS catalog represents more than orders of magnitude increase in the number of well-characterized binary systems compared to pre-\gaia, catalogs \citep{gaiacollaboration2023}. 

Our verification as shown in Fig~\ref{fig:chi2_diff_gnss_dark} demonstrates that the vast majority of sources in the NSS catalog show  preference for binary-star models when fitting their XP spectra, providing independent validation of our methodology. 
This correlation is particularly pronounced for double-lined spectroscopic binaries (SB2 and SB2C types), where the $\chi^2$ improvement for binary models over single-star fits consistently exceeds our detection threshold. 
The stronger signal for SB2 systems is expected, as these binaries typically have more comparable flux ratios between components, making their composite spectral signatures more distinctive. 

As shown in Table~\ref{tab:binary_fractions}, for \gaia\ NSS catalog, our methodology correctly classifies 67\% of astrometric/orbital binaries, 81\% of SB1, and 93\% of SB2 systems as binaries based on our parameter-dependent thresholds in subsection \ref{subsec:threshold}. 
The APOGEE binary sample \citep{el-badry2018}, derived from $\sim$20,000 main-sequence star spectra in APOGEE DR13, includes $\sim$2,500 binaries and $\sim$200 triples identified through spectral fitting with composite models, alongside $\sim$700 velocity-variable systems.
For APOGEE, the correct classification rates are 52\% for SB1 and 85\% for SB2 systems, confirming the robustness of our spectral fitting approach.
Additionally, only 34\% of APOGEE's single-star sample is classified as binaries in our XP fitting, further validating our results, as some of these "single" stars may be long-period q$\approx$1 binaries indistinguishable from single stars in earlier models due to limited parallax precision.

The astrometric-only binaries (Orbital) show a more varied response in our analysis, with detection rates correlating strongly with the magnitude difference between components. 
These patterns across different NSS solution types provide insight into the sensitivity limits and complementary nature of spectrophotometric binary detection compared to spectroscopic and dynamic approaches.
To improve usability, we also provide the cross-matched catalog between our binary fits and the NSS catalog in Appendix~\ref{app:nss}.

\begin{figure}
    \centering
    \includegraphics[width=0.99\linewidth]{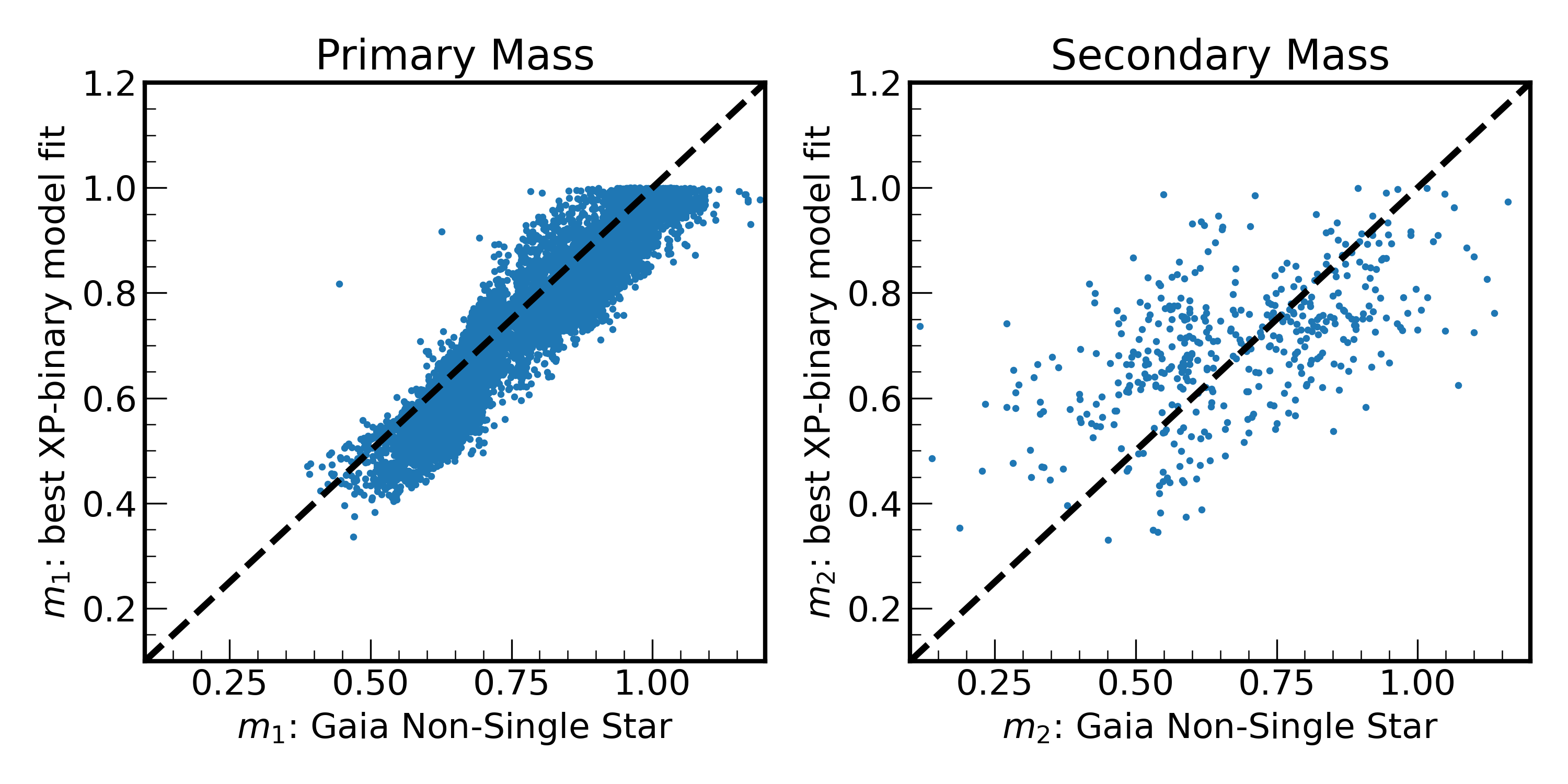}
    \caption{Comparison of primary (left) and secondary (right) masses derived from the XP binary model ($m_1$, $m_2$) against those from the \gaia\ Non-Single Star catalog ($m_1$, $m_2$). 
    The dashed line represents the one-to-one relation.}
    \label{fig:mass_compare}
\end{figure}

We further validate the mass estimates derived from our XP fitting method by comparing them with those provided in the NSS catalog, specifically the \texttt{gaiadr3.binary\_masses} table \citep{gaiacollaboration2023a}. 
This table contains pre-computed masses for the primary ($m_1$) and secondary ($m_2$) components of binary systems. 
The NSS mass estimates are derived using a variety of observational techniques, primarily combining high-precision astrometric, spectroscopic, and, in some cases, photometric (e.g., eclipsing binary) data. 
We adopt the binary mass values from all available solutions, including the solution type, such as `Orbital+SB2' (astrometric orbit combined with double-lined spectroscopy), `EclipsingSpectro(SB2)' (eclipsing binary photometry with SB2 spectroscopy), or `SB1+M1' (single-lined spectroscopy with primary mass from isochrone fitting). 
These methods leverage Kepler’s Third Law and, where applicable, radial velocity semi-amplitudes and orbital inclinations to resolve individual masses.
As shown in Fig.~\ref{fig:mass_compare}, the $m_1$ and $m_2$ values derived from our XP fitting method exhibit strong consistency with the NSS catalog’s mass estimates. 
This agreement validates the robustness of our approach, particularly given the NSS catalog’s reliance on diverse and complementary observational data.

\begin{figure}
    \centering
    \includegraphics[width=1\linewidth]{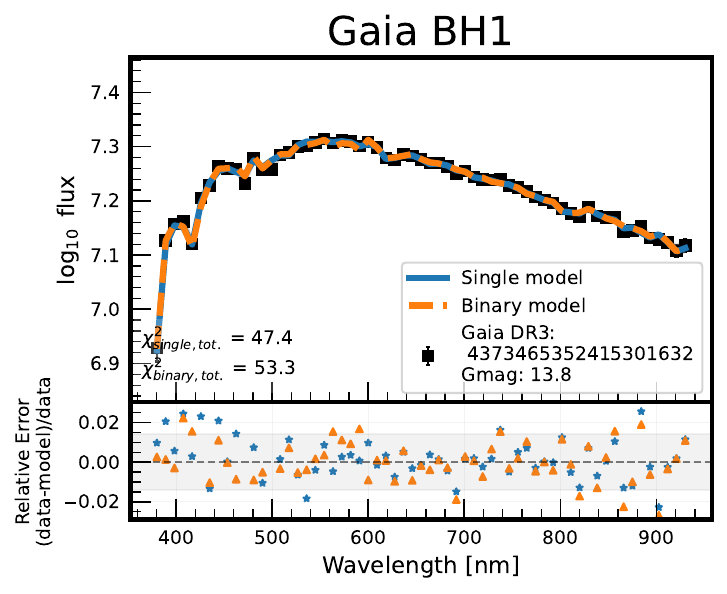}
    \caption{\gaia\ XP spectra and best-fitting models for \gaia\ BH1.
   The black line represents the observed flux.
   The blue solid line shows the best fit single-star model, while the orange dashed line represents the best fit binary model. 
   Bottom panel: Residuals ($\Delta \log \text{flux}$) for both models.}
    \label{fig:bh1}
\end{figure}

\subsection{Validation with Dark Companion Systems}\label{subsec:validation_dark}

The identification of single stars represents the inverse challenge of binary detection. Achieving high purity in binary star identification inevitably comes at the cost of both binary sample completeness and single star sample purity. The chi-squared comparison method requires careful calibration, with threshold settings that vary as demonstrated in Section \ref{subsec:binary_selection}.
To validate our classification approach, we sought benchmark ``single-star-like'' systems to verify whether our method correctly categorizes them. Such validation would be demonstrated by these systems showing minimal improvement in fit quality when modeled as binaries rather than single stars. For this purpose, we utilized two specialized samples:

First, we examined the neutron star companion 
sample from \cite{el-badry2024a}, which contains 21 astrometric binaries with solar-type stars orbiting dark companions with masses near 1.4 M$_\odot$. Since these companions contribute virtually no light to the systems, their spectra should closely resemble those of single stars despite their binary nature.
Second, we analyzed the white dwarf-main sequence (WDMS) binary sample from \cite{li2025}, which identifies approximately 30,000 WDMS systems from XP spectra. In most of these systems, the main sequence primary is at least 10 times brighter than the white dwarf secondary \citep{li2025}, though some were flagged as MSMS binaries using the same chi-squared method. 

The right panel in Figure \ref{fig:chi2_diff_gnss_dark} compares the chi-squared distribution of these samples against a control sample consisting of all \textit{Gaia} sources within 1 kpc. The results demonstrate that for the neutron star sample, the binary model provides minimal improvement over the single-star model ($\log_{10}(\chi^2_\mathrm{MS}) - \log_{10}(\chi^2_\mathrm{MSMS}) < 0.2$). Similarly, 84\% of the WDMS sample shows no notable $\chi^2$ improvement with the binary model, although some systems in this sample could potentially be MSMS binaries with white dwarfs as tertiary companions.
This validation confirms that our method appropriately classifies systems with dark or faint companions as effectively single stars from a spectroscopic perspective, despite their astrometric binary nature.

We also validate the $\chi^2$ difference for \textit{Gaia} Black Hole 1 (BH1) as shown in Figure \ref{fig:bh1}, we see that the total $\chi^2_{\rm single}$ is 47.3, $\chi^2_{\rm binary}$ is 53.3, 
which also proves our method can successfully identify it as a single star / dark companion.
This validation confirms that our method appropriately classifies systems with dark or faint companions as effectively single stars from a spectroscopic perspective.

\begin{figure}
    \centering
    \includegraphics[width=1\linewidth]{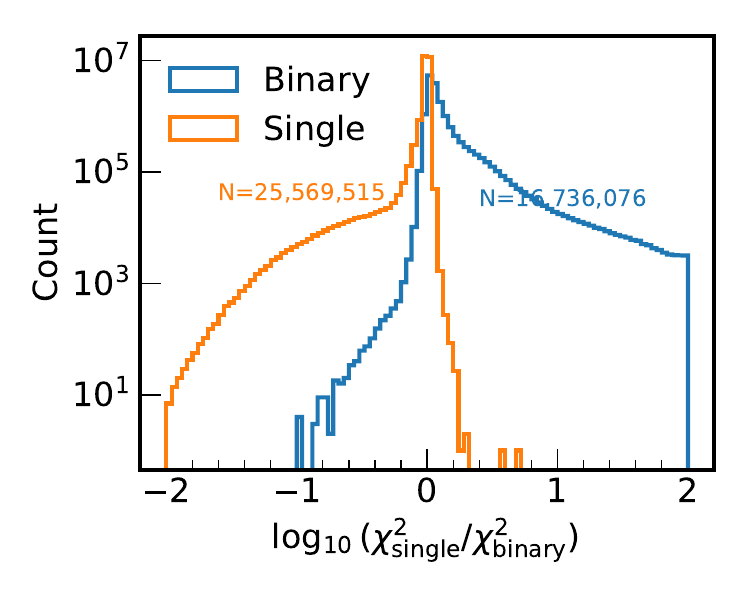}
    \caption{Distribution of \chisqimprove\ for a subset of \gaia\ DR3 XP spectra, illustrating the statistical separation between single-star and binary models. 
    The histogram shows the frequency of sources as a function of $\Delta\chi^2$, with the classifiation based on an optimal threshold $S$ (calibrated to maximize the $F_\beta$-score with $\beta=0.5$) used to classify binary systems. }
    \label{fig:chi2_diff}
\end{figure}

\begin{figure}
    \centering
    \includegraphics[width=0.8\linewidth]{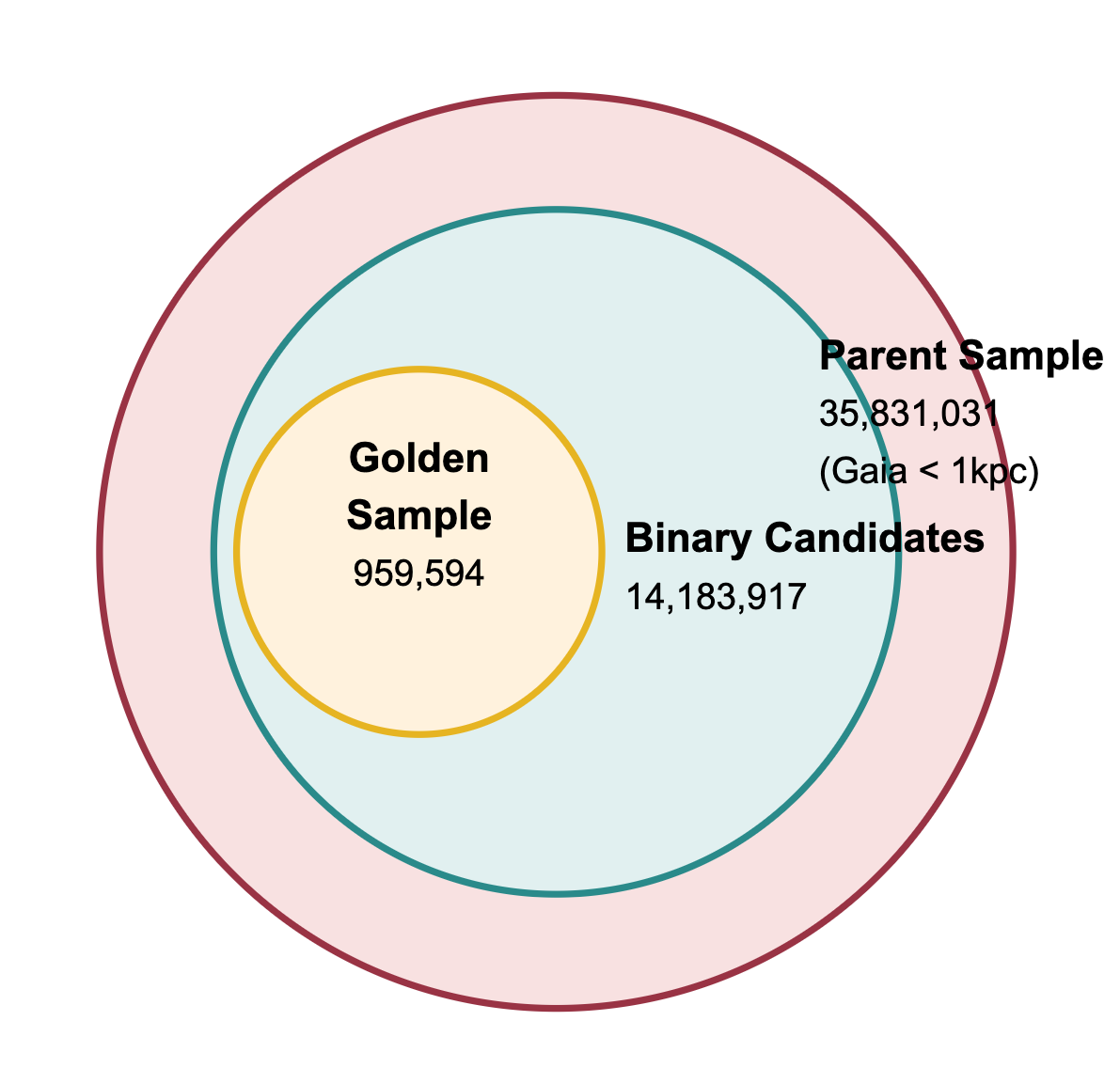}
\caption{Venn diagram showing the hierarchical relationship between our sample selections. 
The parent sample consists of 35 million stars within 1 kpc from Gaia DR3. 
From this population, we identify 14 million binary candidates based on spectral model fitting criteria.
The golden sample is identified through selection criteria in Table \ref{tab:golden_sample}.
}
    \label{fig:venn}
\end{figure}

\section{The MSMS Binary Catalog}\label{sec:binary_catalog}
\subsection{Data Products}

We have compiled and published the main-sequence (MSMS) binary catalog based on \gaia\ DR3 XP spectra. 
This catalog contains 14 million binary candidates identified from our parent sample of 35,831,031 stars within 1 kpc of the Sun. 
The candidates were selected through a combination of spectroscopic modeling techniques that compare the $\chi^2$ values between single-star and binary-star model fits.

For each entry, we include the best-fit results from both single-star and binary spectral models, allowing users to evaluate the binary classification (\ref{fig:venn}). 
The binary model parameters include primary mass, system metallicity, mass ratio, and derived photometric properties such as colors and absolute magnitudes for both components.

The full MSMS Binary Catalog dataset, along with NN weights enabling users to forward model XP single/binary spectra according to their requirements, is now available for public access at \url{https://doi.org/10.5281/zenodo.15166184} \cite{li_2025_15166185}.

\subsection{Binary Candidate Selection}
\label{subsec:binary_selection}
To identify binary systems from our stellar sample, we employ a statistical approach based on the $\chi^2$ difference between single-star and binary-star model fits. 
Our method leverages the model structure described in Section~\ref{sec:method}, extending it to accommodate binary systems by combining the fluxes of two stars with different masses but sharing the same \photmoh, and distance. 
This approach is particularly sensitive to binaries with high mass ratios, which were identified as outliers during the self-cleaning process described in Section~\ref{sec:method}.

We begin with a parent sample of 35 million stars within 1 kpc. 
For each star, we perform two separate model fits: one assuming that the object is a single star and another assuming it is a binary system. 
The chi-square difference ($\Delta\chi^2$) between these two models provides a statistical measure of how much the data favors the binary hypothesis over the single-star hypothesis.

To establish appropriate thresholds for binary classification, we divide our sample into bins based on primary mass (\photmoh) and mass ratio ($q$), as shown in Table~\ref{tab:chi2_thresholds}. 
These thresholds are calibrated to balance completeness and contamination in our binary identification, accounting for the varying sensitivity of our method across different stellar parameters.

A star is classified as a binary candidate if the $\chi^2$ difference exceeds the threshold corresponding to its mass bin:
\begin{equation}
\Delta\chi^2 = \chi^2_{\text{single}} - \chi^2_{\text{binary}} > \text{threshold}(m_1, {\rm [M/H]_{\rm phot}}, q)
\end{equation}

Fig.~\ref{fig:chi2_diff} shows the distribution of $\log_{10}(\chi^2_{\text{single}}/\chi^2_{\text{binary}})$ values for our parent sample. 
Binary candidates systematically exhibit larger chi-squared differences, with the distribution showing a clear bimodality.

Using the thresholds defined in each $m_1$-\photmoh-$q$ bin, we identify 14,183,917 binary candidates from our parent sample, corresponding to an overall binary fraction of approximately 39\%. 
This is consistent with expected binary fractions from previous studies \citep[e.g.][]{gao2014,liu2019, moe2019},  
although the exact fraction varies with stellar mass and metallicity environment.

\begin{figure}
    \centering
\includegraphics[width=0.9\linewidth]{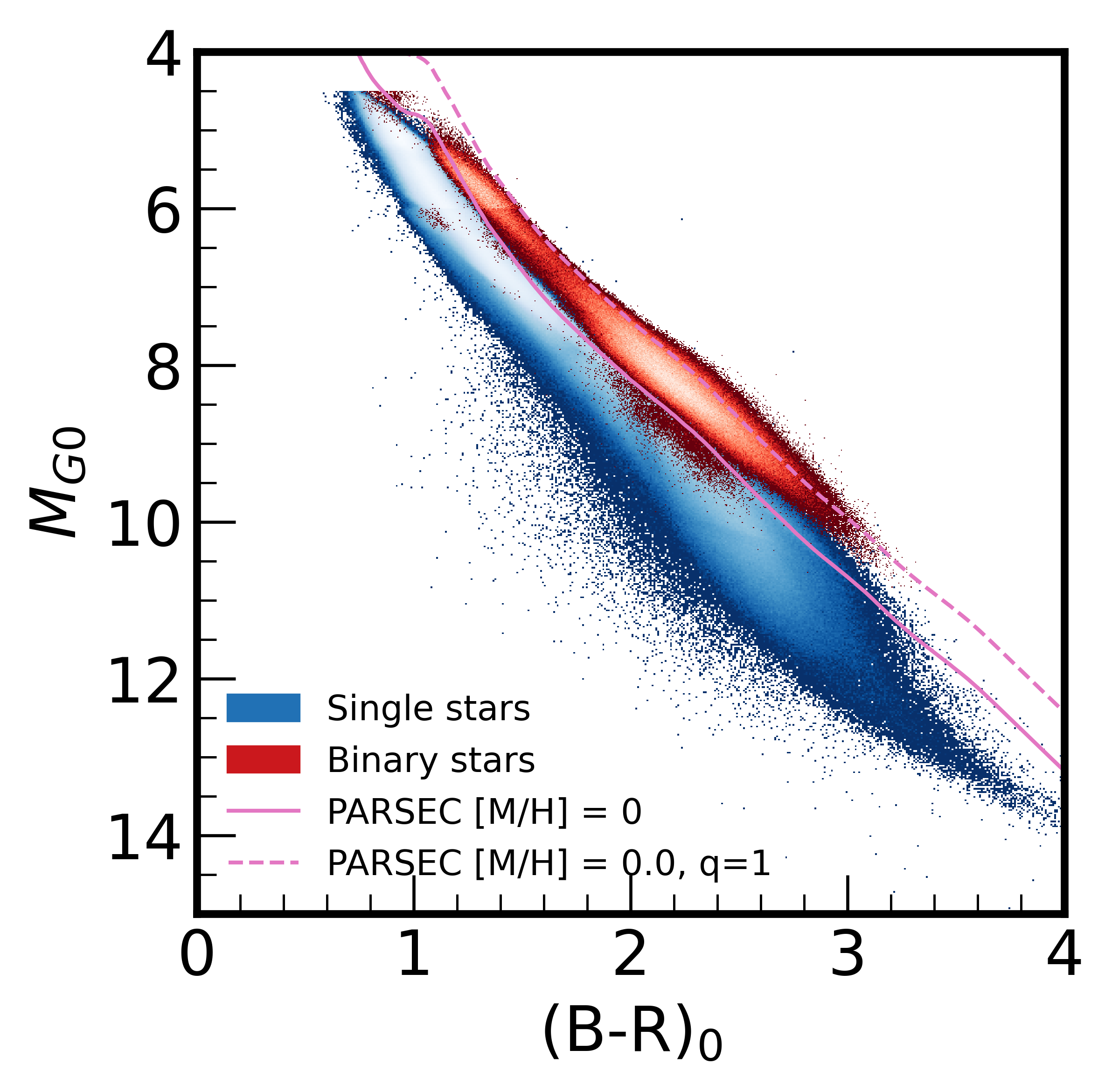}
\caption{\textbf{HR diagrams for golden binary sample of mass-ratio $>0.9$.} 
 The background blue shows the density distribution of selected single stars in, with the scale bar indicating the logarithmic number of stars per bin. 
 The pink solid and dashed lines show PARSEC isochrones \citep{Bressan2012} for [M/H]=0, and $q=1$ binary configurations.
}
    \label{fig:hrd_q_ranges}
\end{figure}

\subsection{Golden Binary Sample}\label{subsec:golden_binary}

\begin{table*}
\centering
\caption{Selection criteria for the golden binary sample}
\label{tab:golden_sample}
\begin{tabular}{l p{8cm}}
\hline\hline
Criterion & Description \\
\hline
predicted\_class = 1 & Binary classification that passes the $\chi^2$ threshold in subsection ~\ref{subsec:binary_selection}\\
$\log_{10}(\chi^2_{\text{single}}/\chi^2_{\text{binary}}) > 0.2$ & $\chi^2$ improvement for binary fit \\
$\chi^2_{\text{single}}/\text{Dof}<3 \lor \chi^2_{\text{binary}}/\text{Dof}<3$ & Good fit quality (either single or binary model) \\
\texttt{fit\_status\_single} $\geq 2 ~\lor$ \texttt{fit\_status\_binary} $\geq 2$ & Successful convergence (either single or binary model) \\
$m_1 < 1.0~M_\odot$ & Primary mass cut \\
\moh > -1 & \photmoh\ cut \\
$M_G \geq 4.5$ & Absolute magnitude cut \\
\texttt{E\_binary} $<0.2$ & Extinction cut  \\
$0.4 \leq q \leq 1.0$  & mass-ratio cut \\
$0.2 \leq \texttt{flux\_ratio\_mean} \leq 1.0$  & flux-ratio cut \\
\texttt{flag\_extrap\_m2}  == False  & secondary mass cut ($m_2 > 0.1$) \\
\hline
\end{tabular}
\end{table*}

To facilitate studies requiring higher confidence binary identifications, we defined a "golden sample" of 959,594 binary systems, representing the most reliable  binary candidates. 
These golden sample members satisfy stringent selection criteria and quality cut detailed in Table ~\ref{tab:golden_sample}. 

The "golden sample" of 959,594 systems, selected from an initial catalog of 35,831,031 systems (\autoref{tab:catalog}), represents the highest confidence binary candidates identified in this work. 
These systems were chosen to ensure robust binary identification and focus on well-characterized main-sequence pairs suitable for detailed analysis. 

These golden sample members satisfy stringent criteria including a significant chi-square improvement with $\log_{10}(\chi^2_{\text{single}}/\chi^2_{\text{binary}}) > 0.2$, good fit quality for both single or binary models with $\chi^2 < 3 \times 61$ (3 per degree of freedom), successful convergence of model fits, primary masses restricted to $M_1 < 1.0~M_\odot$, \photmoh\ from the binary fit $>-1$, absolute G-band magnitude $M_G \geq 4.5$ to focus primarily on main sequence stars, and extinction $E_{\text{binary}} < 0.2$ to ensure the extinction isn't too severe. 
Additionally, systems must have reliable mass ratios between $0.4 \leq q \leq 1.0$, and flux ratios between $0.2$ and $1.0$. 
These cuts are based on practical considerations from mock tests and are not overly stringent, though they may exclude some binaries.

Fig.~\ref{fig:venn} illustrates the relationship between our parent sample, binary candidates, and golden binary sample using a Venn diagram, highlighting the nested nature of these increasingly selective subsets.
Fig.~\ref{fig:hrd_q_ranges} shows HR diagrams for our binary systems across different mass ratio ranges. 
The distinct positions of equal-mass binaries (right panel) compared to lower mass-ratio systems (left panel) in the color-magnitude diagram support the effectiveness of our classification approach, with nearly equal-mass binaries appearing systematically brighter than their single-star counterparts of the same color.
Our catalog includes the stellar parameters of both the primary and the inferred secondary companion for all binary candidates, along with the $\Delta\chi^2$ value and classification confidence metrics.
The MSMS Binary Catalog represents an extensieve uniform samples of binary systems to date, enabling population studies across different Galactic environments and investigating the dependence of binary fraction on stellar and environmental parameters.


\section{Discussion and Conclusions}\label{sec:discussion}

We discuss binary star systems identified from \textit{Gaia} DR3 XP spectra in the following sections. 
In Section~\ref{subsec:parallax_error}, we first validate our spectral fitting results with enhanced parallax error treatment, demonstrating the robustness of our methodology.
We then examine potential caveats and limitations of our binary analysis in Section~\ref{subsec:caveats}.

\subsection{Parallax Errors}\label{subsec:parallax_error}

\begin{figure}
    \centering
    \includegraphics[width=1\linewidth]{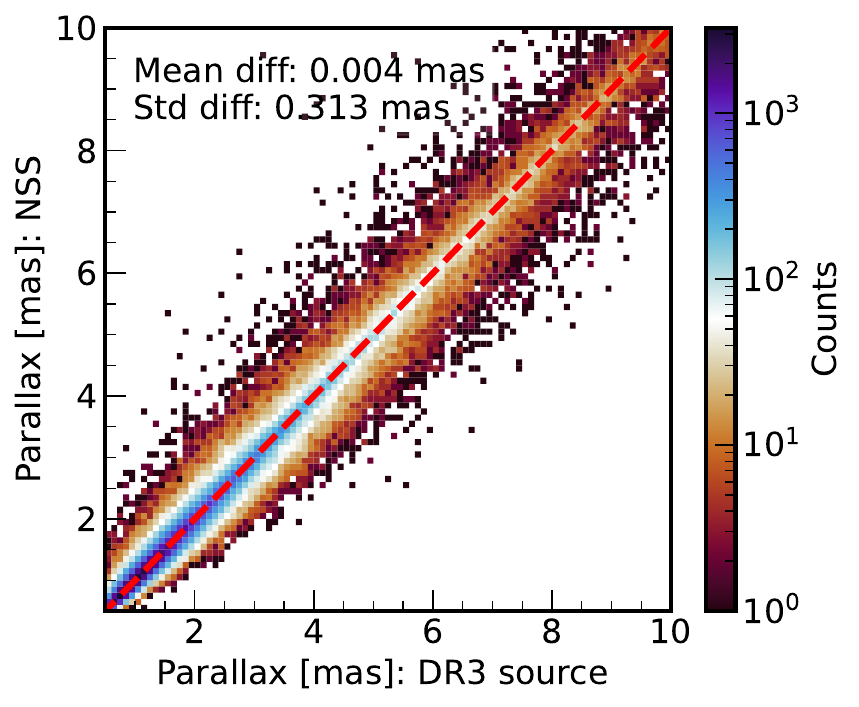}
    \caption{Comparison of parallax from \gaia\ DR3 source catalog and NSS calalog. The dashed red line represents the one-to-one relation, and the color scale denotes the number density.}
    \label{fig:gnss_parallax}
\end{figure}

\begin{figure*}
    \centering
    \includegraphics[width=0.48\linewidth]{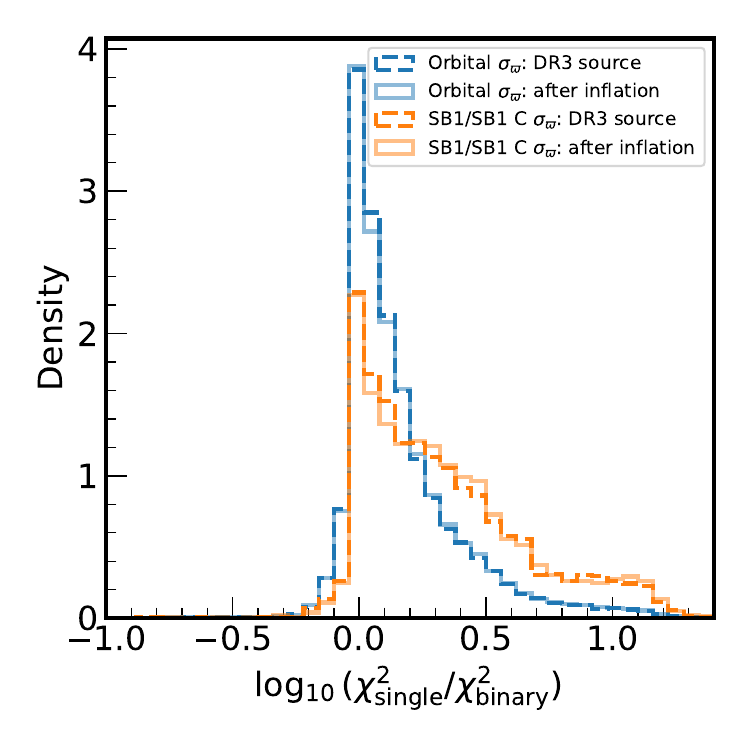}
    \includegraphics[width=0.48\linewidth]{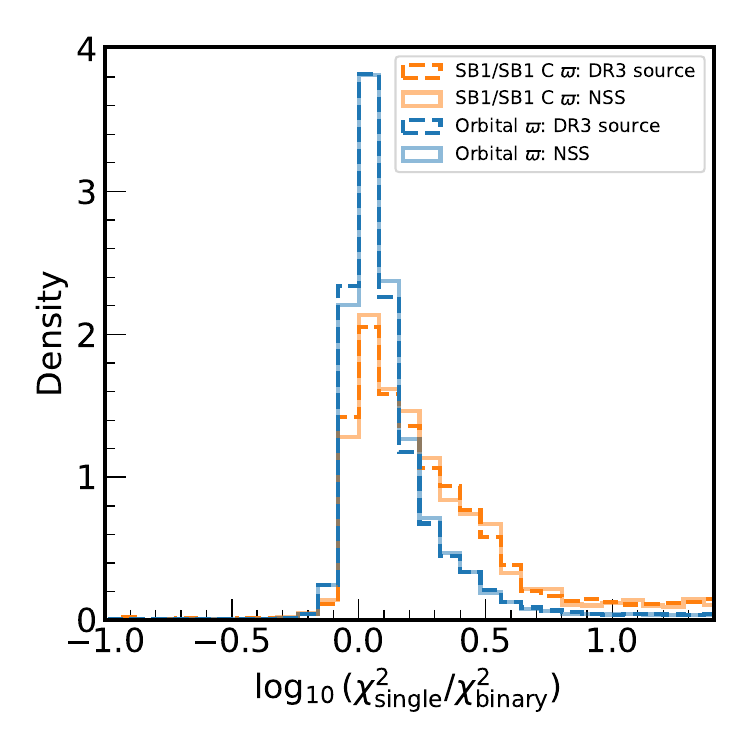}
    \caption{Distribution of the logarithmic $\chi^2$ improvement for binary models. In each panel, blue lines denote astrometric binary stars classified by the NSS catalog, while orange lines represent SB1 binaries.
    Left: Dashed lines are derived from the original DR3 source catalog parallax errors, while solid lines show results after error inflation following the model of \citet{elbadry2025usegaiaparallaxesstars}.
    Right: The dashed lines use parallaxes from the DR3 source catalog, while solid lines use parallaxes from the NSS catalog, which incorporate binary orbital solutions in their parallax estimates.}
    \label{fig:chi2_diff_plxerr}
\end{figure*}

Two factors contribute to the underestimation of uncertainties for true binaries in our analysis. 
First, treating binaries as single stars introduces errors in parallax estimates \citep{gaiacollaboration2023a}. We validate the fitting results using parallax solutions derived from binary orbit assumptions.
Second, binaries exhibit poor astrometric fits due to orbital motion, leading to underestimated uncertainties in Gaia’s astrometric solutions, including parallax. 
We apply a simulation-driven method \citep{elbadry2025usegaiaparallaxesstars} to correct the parallax error and compare the results with our findings.

\subsubsection{Comparison with Parallax from Binary Solutions}

The NSS catalog fits complex models to describe the behavior of non-single stars, accounting for orbital motion in binary systems \citep{gaiacollaboration2023a}. 
It includes solutions for astrometric, spectroscopic, and eclipsing binaries, providing parallax estimates derived from these models. 
The comparison between DR3 source and NSS parallax solutions (Fig. ~\ref{fig:gnss_parallax}) reveals a mean difference of 0.004 mas with an RMS scatter of 0.3 mas, indicating that source parallaxes slightly underestimate the true values of the NSS binary systems. 
We  further compare the $\chi^2$ improvement using different parallax sources in Figure~\ref{fig:chi2_diff_plxerr}, where the right panel demonstrates that the difference between NSS parallax and source parallax is negligible. 
This suggests that for sources without NSS orbital solutions, using the DR3 source parallax for XP fitting may be a reasonable approach.

\subsubsection{Inflating Parallax Uncertainties}

To address underestimation, we adopt the inflation model from \citet{elbadry2025usegaiaparallaxesstars} to adjust uncertainties based on the Renormalized Unit Weight Error (RUWE), a metric of astrometric fit quality where higher values indicate poorer fits. 
Many binaries with close companions have elevated RUWE values due to photocenter oscillations from orbital motion, which single-star models cannot accommodate \citep{gaiacollaboration2023a}. 
For binaries with orbital periods of a few years, photocentric acceleration further increases RUWE. 
The model of \citet{elbadry2025usegaiaparallaxesstars} adjusts reported uncertainties as a function of RUWE, parallax, and apparent magnitude, transforming error distributions into Gaussian forms.

We compare the $\chi^2$ improvement (\chisqimprove) in Figure~\ref{fig:chi2_diff_plxerr}.
For astrometric (orbital) binaries, the difference before and after the parallax inflation is negligible. 
However, for SB1 systems, we observe modest differences: 
For binaries with \chisqimprove$<0.5$, the XP fit incorporating parallax error inflation demonstrates better performance than without inflation. Specifically, binaries previously showing \chisqimprove\ between 0 and 0.2 shift to the 0.2-0.5 bin after applying parallax error inflation, indicating improved classification capability.
For binaries with \chisqimprove$>0.5$, parallax error inflation produces no significant difference in results.

\subsection{Caveats and Limitations}\label{subsec:caveats}

\subsubsection{Extinction Modeling}
Our extinction modeling relies on an average extinction curve derived from Gaia XP spectra \citep{zhang2023a}. 
This approach, while effective for population-level analysis, may not fully capture the variation in extinction properties across different Galactic environments. 
Recent studies by \citet{Zhang2025} demonstrate that $R(V)$, which characterizes the slope of the extinction curve in optical wavelengths and is related to the typical size of dust grains, can vary depending on the line of sight. 
This variation could introduce systematic uncertainties in our derived stellar parameters, particularly in regions with anomalous dust properties or high extinction values. 
Future work could incorporate spatially-dependent extinction curves to further refine the binary parameter estimates.

\subsubsection{Low-Mass Secondary Extrapolation}
For binary systems with secondary masses ($m_1 \times q$) below 0.1 M$_\odot$, our parameter estimates should be treated with caution. 
The stellar structure models we employ \citep{Bressan2012} have a lower mass limit of 0.1 M$_\odot$, meaning that binary fits with low-mass secondaries rely on extrapolation from our NN emulators. 
We provide a quality flag \texttt{flag\_m2\_extrap} in our catalog to identify these systems. 

\subsubsection{Evolutionary Effects for Higher-Mass Stars}
For stars more massive than approximately 0.8 M$_\odot$, which corresponds to the highest mass that remains on the main sequence after 12.5 Gyr at [Fe/H] = -2 \citep{Paxton2011, Choi2016}, evolutionary effects become increasingly important. 
These stars likely have evolved from their zero-age main sequence positions in the color-magnitude diagram, making our modeling from mass and metallicity to color-absolute magnitude less accurate without accounting for age variations. 
Current parameter estimates for these higher-mass systems do not incorporate age as a variable, potentially introducing systematic biases in the inferred mass ratios and metallicities. 
A future refinement of this work should consider stellar ages and initial masses of binary systems when modeling their spectral and photometric properties.

\subsection{Summary}

This study presents the main-sequence binary (MSMS) Catalog, identifying $\sim$14 million binary star candidates from Gaia Data Release 3 (DR3) BP/RP spectra. The analysis uses spectrophotometric data for 35 million stars within 1 kpc of the Sun, selected from 220 million Gaia DR3 sources. 

Binaries are identified when the binary model yields a lower $\chi^2$ than the single-star model, with a parameter-dependent $\chi^2$ improvement threshold for mass ratios $0.5 \leq q \leq 1.0$. 
A subset of approximately one million systems (``golden sample") is defined using a stricter chi-squared improvement, achieving 93\% completeness of SB2s validated using the \gaia\ non-single star catalog.

This approach overcomes limitations of spectroscopic methods, which rely on radial velocity shifts and favor short-period binaries (periods < 1000 days). 
By comparing spectral fits, the method detects binaries across mass ratios $0.5 \leq q \leq 1.0$. Validation with synthetic spectra confirms recovery of primary mass and\photmoh with errors < 10\%. 
The 14 million candidates reveals a binary fraction of 39\% among main-sequence stars, with mass- and metallicity-dependent trends in the mass-ratio distribution. 

\begin{acknowledgements}
JL thanks Coryn Bailer-Jones for constructive discussion and insights.
We acknowledge support from the European Research Council through ERC Advanced Grant No. 101054731.
This work has made use of data from the European Space Agency (ESA) mission {\it Gaia} (\url{https://www.cosmos.esa.int/gaia}), processed by the {\it Gaia} Data Processing and Analysis Consortium (DPAC, \url{https://www.cosmos.esa.int/web/gaia/dpac/consortium}). Funding for the DPAC has been provided by national institutions, in particular the institutions participating in the {\it Gaia} Multilateral Agreement. YST is supported by the National Science Foundation under Grant AST-2406729.
\end{acknowledgements}

\bibliographystyle{aa} 
\bibliography{MSMS}

\begin{appendix}
\begin{figure}
    \centering
    \includegraphics[width=0.99\linewidth]{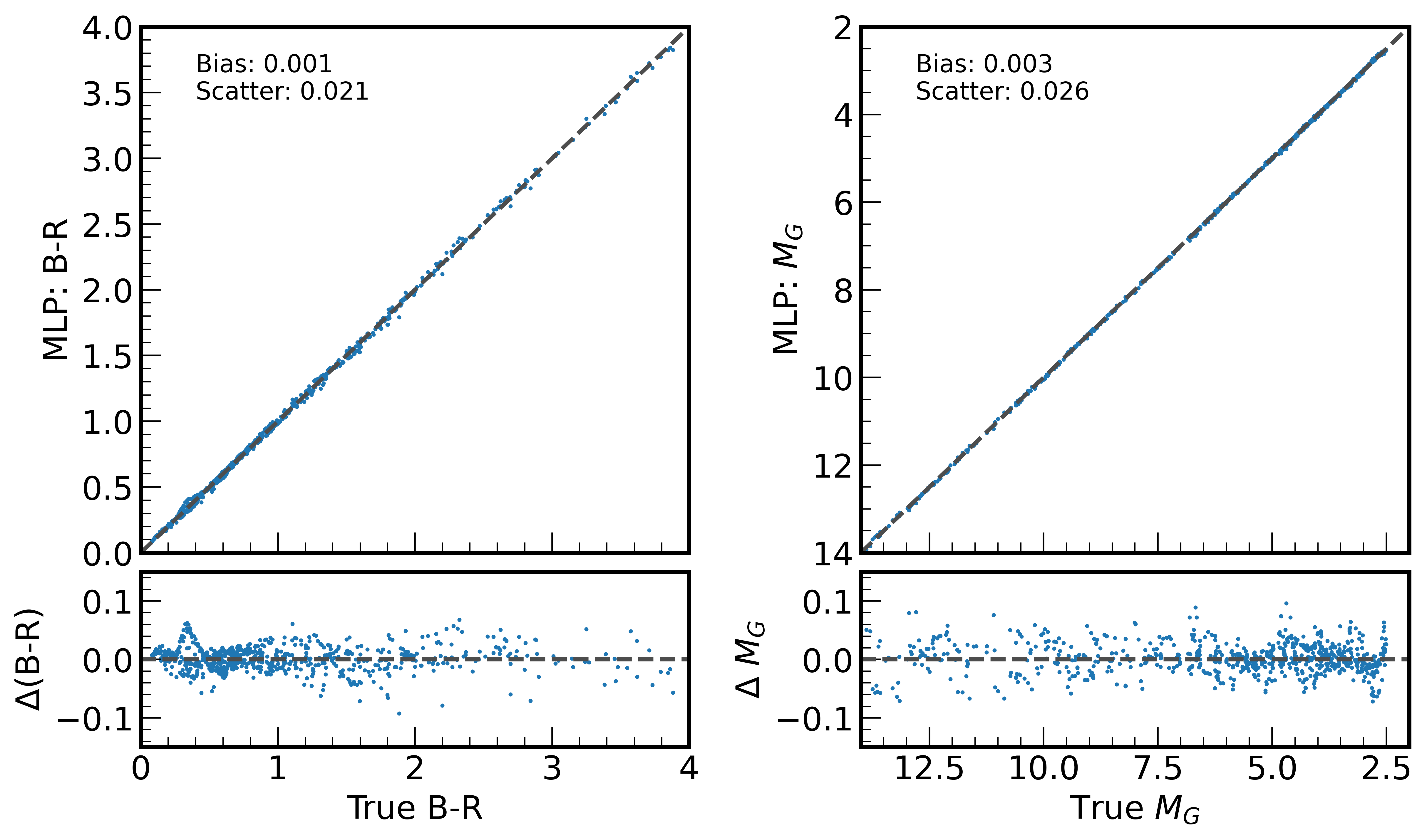}
    \caption{Validation of the neural network-based mapping function $\mathcal{M}$ that translates stellar physical parameters to H-R diagram coordinates. 
    Left panels: Comparison between true and predicted B-R color, showing good agreement with minimal bias (0.001) and scatter (0.021). 
    Right panels: Validation of absolute G-magnitude ($M_G$) predictions, displaying negligible bias (0.003) and low scatter (0.026). 
    The bottom panels show residuals (predicted minus true values) across the parameter range, demonstrating no systematic trends.}
    \label{fig:compare_mlp}
\end{figure}

\section{Sampled XP spectra}\label{app:sample_xp}

To accurately process XP spectra, we must account for covariances between observed fluxes across wavelengths. 
Our sampling range of 392--992~nm and step size of 10~nm were chosen to trim noisy spectral edges and ensure positive-definite covariance matrices \citep{zhang2023a}.

The $\chi^2$ statistic between predicted and observed fluxes is calculated as:
\begin{equation}
\chi^2 = \Delta \vec{f}^T C_{\vec{f}_{\rm obs}}^{-1} \Delta \vec{f} = \left| L \Delta \vec{f} \right|^2,
\end{equation}
where $\Delta \vec{f} \equiv \vec{f}_{\rm pred} - \vec{f}_{\rm obs}$ represents the residuals between predicted and observed fluxes.

To ensure numerical stability in this calculation, we diagonalize each covariance matrix using its eigendecomposition:
\begin{equation}
C_{\vec{f}_{\rm obs}} = U D U^T,
\end{equation}
where $U$ is orthonormal and $D$ is diagonal. We impose a minimum value of $10^{-9}$ (in units of $10^{-36}\,\mathrm{W^2\,m^{-4}\,nm^{-2}}$) on the diagonal elements of $D$ (yielding $\hat{D}$) and define $L \equiv \hat{D}^{-1/2} U^T$.

The covariance matrix $\mathbf{C}_{\vec{f}_{\text{obs}}}$ incorporates several uncertainty components:
\begin{equation}
\begin{split}
\mathbf{C}_{\vec{f}_{\text{obs}}} = \mathbf{C}_{\vec{f}_{\text{gaia}}} \! + \, \text{diag}\left(0.005 \vec{f}_{\text{gaia}}\right)^{\! 2} \\
+ 0.005^2 \vec{f}_{\text{gaia}} \vec{f}_{\text{gaia}}^T + 0.001^2 \vec{f}_{\text{BP}} \vec{f}_{\text{BP}}^T + 0.001^2 \vec{f}_{\text{RP}} \vec{f}_{\text{RP}}^T,
\end{split}
\end{equation}
where $\vec{f}_{\rm gaia}$ represents the sampled flux, and $\vec{f}_{BP}$ and $\vec{f}_{RP}$ are zero-padded vectors containing the sampled BP and RP spectra, respectively.

The terms in this expression account for:
\begin{enumerate}
    \item The sample-space covariance matrix from \gaia\ DR3, denoted as $\mathbf{C}_{\vec{f}_{\text{gaia}}}$
    \item A 0.5\% independent flux uncertainty at each wavelength
    \item A 0.5\% uncertainty in the zero-point calibration of the overall spectral flux
    \item A 0.1\% uncertainty in the zero-point calibration of the BP spectrum
    \item A 0.1\% uncertainty in the zero-point calibration of the RP spectrum
\end{enumerate}

Following \cite{zhang2023a}, we inflate the original covariance matrix to avoid dominance by small uncertainties reported in \gaia\ DR3. 
The zero-point correction involves three components: the overall spectral zero-point and the relative zero-points for BP and RP spectra.

\section{From physical parameters to CMD}\label{app:hr_mapping}

The mapping function $\mathcal{M}$ that translates stellar physical parameters to H-R diagram coordinates is implemented as a neural network with three hidden layers, maintaining consistency with the spectral emulator architecture described in Section~\ref{subsec:spectral_model}.

Figure \ref{fig:compare_mlp} validates our stellar parameter-to-observable mapping function $\mathcal{M}$, showing notable accuracy in translating physical parameters to H-R diagram positions. 
The correlation between true and predicted color indices (B-R) shows high precision with minimal bias (0.001) and scatter (0.021), demonstrating that our NN implementation accurately captures the nonlinear relations between mass, \photmoh, and color. 
Similarly, absolute magnitude predictions ($M_G$) exhibit agreement with true values, with negligible bias (0.003) and low scatter (0.026). 
The residual plots (bottom panels) show no notable systematic trends across the parameter space, confirming that our implementation of the PARSEC evolutionary tracks through the NN mapping function successfully reproduces the photometric properties of stars with diverse masses and metallicities.
This approach ensures that the physical parameters inferred from our spectral analysis can be converted to observable quantities for comparison with photometric data and for visualization on standard H-R diagrams.

\section{The Catalog}

\subsection{Binary candidate selection}
\begin{table*}[ht]
\centering
\caption{Binary Classification Thresholds and Performance Metrics by Parameter Bins}
\label{tab:chi2_thresholds}
\begin{tabular}{ccccccc}
\hline
\multicolumn{3}{c}{Parameter Bins} & \multirow{2}{*}{Threshold} & \multicolumn{3}{c}{Performance Metrics} \\
\cline{1-3} \cline{5-7}
Stellar Mass & \photmoh & Mass Ratio $q$ & & $F_{\beta}$ & Purity & Completeness \\
\hline
$0.1$-$0.5$ & $-2.0$-$-1.0$ & $0.1$-$0.5$ & $4.75$ & $80.1\%$ & $96.0\%$ & $48.1\%$ \\
$0.1$-$0.5$ & $-2.0$-$-1.0$ & $0.5$-$0.9$ & $1.27$ & $92.5\%$ & $95.5\%$ & $82.2\%$ \\
$0.1$-$0.5$ & $-2.0$-$-1.0$ & $0.9$-$1.0$ & $-0.77$ & $88.8\%$ & $88.1\%$ & $91.6\%$ \\
$0.1$-$0.5$ & $-1.0$-$-0.5$ & $0.1$-$0.5$ & $5.09$ & $73.3\%$ & $95.8\%$ & $37.8\%$ \\
$0.1$-$0.5$ & $-1.0$-$-0.5$ & $0.5$-$0.9$ & $1.61$ & $92.8\%$ & $95.7\%$ & $82.9\%$ \\
$0.1$-$0.5$ & $-1.0$-$-0.5$ & $0.9$-$1.0$ & $-1.91$ & $83.5\%$ & $80.8\%$ & $96.3\%$ \\
$0.1$-$0.5$ & $-0.5$-$0.0$ & $0.1$-$0.5$ & $4.42$ & $78.5\%$ & $94.3\%$ & $47.0\%$ \\
$0.1$-$0.5$ & $-0.5$-$0.0$ & $0.5$-$0.9$ & $2.12$ & $93.8\%$ & $95.8\%$ & $86.6\%$ \\
$0.1$-$0.5$ & $-0.5$-$0.0$ & $0.9$-$1.0$ & $3.73$ & $87.0\%$ & $97.3\%$ & $61.2\%$ \\
$0.1$-$0.5$ & $0.0$-$0.5$ & $0.1$-$0.5$ & $5.11$ & $85.3\%$ & $99.2\%$ & $54.6\%$ \\
$0.1$-$0.5$ & $0.0$-$0.5$ & $0.5$-$0.9$ & $5.35$ & $96.8\%$ & $99.3\%$ & $87.9\%$ \\
$0.1$-$0.5$ & $0.0$-$0.5$ & $0.9$-$1.0$ & $6.11$ & $97.0\%$ & $99.4\%$ & $88.4\%$ \\
$0.5$-$1.0$ & $-2.0$-$-1.0$ & $0.1$-$0.5$ & $3.47$ & $72.3\%$ & $83.8\%$ & $46.6\%$ \\
$0.5$-$1.0$ & $-2.0$-$-1.0$ & $0.5$-$0.9$ & $1.82$ & $87.6\%$ & $92.2\%$ & $73.2\%$ \\
$0.5$-$1.0$ & $-2.0$-$-1.0$ & $0.9$-$1.0$ & $-$ & $-$ & $-$ & $-$ \\
$0.5$-$1.0$ & $-1.0$-$-0.5$ & $0.1$-$0.5$ & $3.54$ & $74.1\%$ & $89.8\%$ & $43.6\%$ \\
$0.5$-$1.0$ & $-1.0$-$-0.5$ & $0.5$-$0.9$ & $4.04$ & $89.2\%$ & $96.8\%$ & $68.0\%$ \\
$0.5$-$1.0$ & $-1.0$-$-0.5$ & $0.9$-$1.0$ & $-7.11$ & $80.4\%$ & $76.7\%$ & $99.6\%$ \\
$0.5$-$1.0$ & $-0.5$-$0.0$ & $0.1$-$0.5$ & $3.80$ & $72.2\%$ & $91.3\%$ & $39.4\%$ \\
$0.5$-$1.0$ & $-0.5$-$0.0$ & $0.5$-$0.9$ & $2.32$ & $89.7\%$ & $94.1\%$ & $75.7\%$ \\
$0.5$-$1.0$ & $-0.5$-$0.0$ & $0.9$-$1.0$ & $2.95$ & $74.9\%$ & $89.3\%$ & $45.5\%$ \\
$0.5$-$1.0$ & $0.0$-$0.5$ & $0.1$-$0.5$ & $3.92$ & $77.8\%$ & $92.5\%$ & $47.6\%$ \\
$0.5$-$1.0$ & $0.0$-$0.5$ & $0.5$-$0.9$ & $4.40$ & $94.7\%$ & $97.9\%$ & $83.7\%$ \\
$0.5$-$1.0$ & $0.0$-$0.5$ & $0.9$-$1.0$ & $8.29$ & $97.7\%$ & $99.7\%$ & $90.2\%$ \\
$1.0$-$1.5$ & $-2.0$-$-1.0$ & $0.1$-$0.5$ & $3.12$ & $55.4\%$ & $67.0\%$ & $32.8\%$ \\
$1.0$-$1.5$ & $-2.0$-$-1.0$ & $0.5$-$0.9$ & $3.86$ & $89.5\%$ & $95.4\%$ & $71.7\%$ \\
$1.0$-$1.5$ & $-2.0$-$-1.0$ & $0.9$-$1.0$ & $4.58$ & $96.2\%$ & $98.3\%$ & $88.4\%$ \\
$1.0$-$1.5$ & $-1.0$-$-0.5$ & $0.1$-$0.5$ & $2.03$ & $56.8\%$ & $66.8\%$ & $35.5\%$ \\
$1.0$-$1.5$ & $-1.0$-$-0.5$ & $0.5$-$0.9$ & $4.32$ & $89.7\%$ & $95.5\%$ & $72.4\%$ \\
$1.0$-$1.5$ & $-1.0$-$-0.5$ & $0.9$-$1.0$ & $45.54$ & $91.6\%$ & $100.0\%$ & $68.5\%$ \\
$1.0$-$1.5$ & $-0.5$-$0.0$ & $0.1$-$0.5$ & $1.90$ & $63.3\%$ & $76.0\%$ & $37.9\%$ \\
$1.0$-$1.5$ & $-0.5$-$0.0$ & $0.5$-$0.9$ & $1.90$ & $87.8\%$ & $92.8\%$ & $72.2\%$ \\
$1.0$-$1.5$ & $-0.5$-$0.0$ & $0.9$-$1.0$ & $4.41$ & $85.9\%$ & $96.7\%$ & $59.4\%$ \\
$1.0$-$1.5$ & $0.0$-$0.5$ & $0.1$-$0.5$ & $2.91$ & $63.4\%$ & $84.9\%$ & $31.5\%$ \\
$1.0$-$1.5$ & $0.0$-$0.5$ & $0.5$-$0.9$ & $3.38$ & $89.1\%$ & $96.9\%$ & $67.6\%$ \\
$1.0$-$1.5$ & $0.0$-$0.5$ & $0.9$-$1.0$ & $3.34$ & $82.9\%$ & $90.9\%$ & $61.2\%$ \\
\hline
\multicolumn{3}{l}{Overall} & $-$ & $83.6\%$ & $91.6\%$ & $66.0\%$ \\
\hline
\end{tabular}
\end{table*}

We provide both the single and binary fitting results for $\sim 35$ million sources in \gaia\ database.
For each source, we include the optimal parameters from both single-star and binary-star models, along with quality flags and derived quantities. 
The catalog contains 14,183,917 binary candidates, with a subset of 959,594 sources meeting our stricter criteria for the "golden sample." 
For each source, we provide the primary and secondary stellar parameters (masses, metallicities), the mass ratio, flux ratios in different bands, and quality metrics including $\chi^2$ values and fit convergence flags. 
The catalog also includes \gaia\ parallax and extinction estimates from both our model fits and external dust maps. Users can filter the catalog based on the provided quality flags to select samples optimized for either completeness or purity depending on their scientific objectives.

\begin{table*}\centering
\caption{Parameter Descriptions for the MSMS Binary Catalog (Parent Sample, Binary Candidates, and Golden Sample)}
\label{tab:catalog}
\begin{tabular}{llp{10cm}}\hline\hline
Parameter & Type & Description \\\hline
\texttt{source\_id} & Int & Gaia DR3 source identifier \\\hline
\multicolumn{3}{c}{\textbf{Fundamental parameters}} \\\hline
\texttt{ra} & Double & Right ascension by \gaia\ DR3 (deg) \\
\texttt{dec} & Double & Declination by \gaia\ DR3 (deg) \\
\texttt{bp\_rp} & Double & Gaia DR3 B-R colors \\
\texttt{phot\_g\_mean\_mag} & Double & Gaia DR3 G-band magnitude \\
\texttt{parallax} & Double & Gaia DR3 parallax (mas) \\
\texttt{parallax\_error} & Double & Uncertainty in parallax measurement \\
\texttt{radial\_velocity} & Double & Gaia DR3 radial velocity (km/s) \\
\texttt{radial\_velocity\_error} & Double & Uncertainty in Gaia DR3 radial velocity (km/s) \\
\texttt{E\_edhf} & Double & Extinction from Bayestar19 dust map \\
\texttt{A\_G\_edhf}, \texttt{A\_B\_edhf}, \texttt{A\_R\_edhf} & Double & Extinction in G, B, and R bands from \cite{Edenhofer2024}'s dustmap \\
\hline
\multicolumn{3}{c}{\textbf{Best-fit results for single XP spectral model}} \\\hline
\texttt{mass\_single} & Double & Stellar mass from single-model (M$_\odot$)\\
\texttt{e\_mass\_single} & Double & Uncertainty in stellar mass from single-model (M$_\odot$)\\
\texttt{mh\_single} & Double & Metallicity from single-model solution [M/H] \\
\texttt{e\_mh\_single} & Double & Uncertainty in metallicity from single-model solution \\
\texttt{E\_single} & Double & Extinction from single-model\\
\texttt{e\_E\_single} & Double & Uncertainty in extinction from single-model\\
\texttt{chi2\_mass\_mh}& Double & $\chi^2$ value of the single-model fit \\\hline

\multicolumn{3}{c}{\textbf{Best-fit results for binary XP spectral model}} \\\hline
\texttt{m1\_binary} & Double & Primary star mass (M$_\odot$) \\
\texttt{e\_m1\_binary} & Double & Uncertainty in primary star mass (M$_\odot$) \\
\texttt{m2\_binary} & Double & Secondary star mass (M$_\odot$) \\
\texttt{mh\_binary} & Double & System photometric metallicity [M/H]\\
\texttt{e\_mh\_binary} & Double & Uncertainty in system photometric metallicity \\
\texttt{q\_binary} & Double & Mass ratio (m$_2$/m$_1$) \\
\texttt{e\_q\_binary} & Double & Uncertainty in mass ratio \\
\texttt{E\_binary} & Double & Extinction from binary-model\\
\texttt{e\_E\_binary} & Double & Uncertainty in extinction from binary-model\\
\texttt{chi2\_mass\_mh\_binary}& Double & $\chi^2$ value of the binary-model fit \\
\texttt{flux\_ratio\_mean} & Double & Mean XP flux ratio between primary and secondary components \\
\texttt{flux\_ratio\_std} & Double & Standard deviation of the XP flux ratio between primary and secondary components \\
\texttt{b\_r\_prim} & Double & B-R color of the primary component\\
\texttt{b\_r\_second} & Double & B-R color of the secondary component\\
\texttt{MG\_prim} & Double & Absolute G magnitude of the primary component \\
\texttt{MG\_second} & Double & Absolute G magnitude of the secondary component \\ \hline

\multicolumn{3}{c}{\textbf{Quality flags}} \\\hline
\texttt{fit\_status\_single}  & int & Status code for single-model fit convergence \\
\texttt{fit\_status\_binary}  & int & Status code for binary-model fit convergence \\
\texttt{fit\_success\_single} & Boolean & Whether the single-model fit was successful \\
\texttt{fit\_success\_binary} & Boolean & Whether the binary-model fit was successful \\
\texttt{flag\_extrap\_m2}     & Boolean & Whether secondary mass from the binary fit lower than 0.1 M$_\odot$\\
\\\hline\hline
\end{tabular}
\end{table*}

\subsection{Quality flags}

The catalog includes \textbf{quality flags} to assess the reliability and convergence status of single- and binary-model fits. 
The \texttt{fit\_status\_single} and \texttt{fit\_status\_binary} codes indicate termination reasons: \texttt{-1} for invalid inputs, \texttt{0} for exceeding function evaluations, and \texttt{1}--\texttt{4} for convergence via \texttt{gtol}, \texttt{ftol}, \texttt{xtol}, or multiple tolerances. 
The Boolean \texttt{fit\_success\_single} and \texttt{fit\_success\_binary} flags denote overall success (typically \texttt{True} for status codes $\geq 1$, subject to additional criteria like physically plausible parameters). 
The \texttt{flag\_extrap\_m2} highlights cases where the binary fit yields a secondary mass $<0.1\,M_\odot$, signaling potential extrapolation beyond the model’s calibrated range. 
These flags collectively help users gauge the robustness of fitted parameters and identify fits requiring scrutiny.

The \texttt{least\_squares} optimizer in \texttt{Scipy} uses three key tolerance parameters to determine convergence:
\begin{itemize}
    \item[\bfseries gtol] \textbf{Gradient tolerance} \\
    Optimization stops when the maximum gradient magnitude falls below this threshold. \\
    Corresponds to quality flag: \texttt{fit\_status\_* = 1}

    \item[\bfseries ftol] \textbf{Function tolerance} \\
    Stops when relative improvements in the cost function become negligible. \\
    Corresponds to quality flag: \texttt{fit\_status\_* = 2}

    \item[\bfseries xtol] \textbf{Parameter tolerance} \\
    Terminates when parameter adjustments become sufficiently small. \\
    Corresponds to quality flag: \texttt{fit\_status\_* = 3} \\

    \item[\bfseries both] \texttt{ftol + xtol} \\
    Stopping condition met for both function and parameter tolerances. \\
    Corresponds to quality flag: \texttt{fit\_status\_* = 4}
\end{itemize}

\noindent Typical default values are $10^{-8}$ for all tolerances. Lower values enforce stricter convergence.


\begin{table*}[htbp]
\centering
\caption{Binary Catalog Information}
\label{tab:catalog_info}
\begin{subtable}{\textwidth}
\centering
\caption{Selected Parameters from the Gaia DR3 Non-Single Star (NSS) Catalog}
\label{tab:nss_params}
\begin{tabular}{lp{8cm}}
\hline
\textbf{Parameter} & \textbf{Description} \\
source\_id & Gaia DR3 source identifier\\
nss\_solution\_type & Binary solution type\\
\hline
\multicolumn{2}{c}{\textit{Radial Velocity Parameters}} \\
\hline
radial\_velocity & Gaia radial velocity (km/s) \\
radial\_velocity\_error & Uncertainty in the radial velocity measurement \\
rv\_chisq\_pvalue & p-value of the chi-square test for radial velocity variability \\
rv\_expected\_sig\_to\_noise & Expected signal-to-noise ratio for radial velocity measurements \\
rv\_nb\_transits & Number of radial velocity transits \\
rv\_renormalised\_gof & Renormalised goodness of fit for radial velocity solutions \\
\hline
\multicolumn{2}{c}{\textit{Orbital Parameters}} \\
\hline
period & Orbital period of the binary system (days) \\
eccentricity & Orbital eccentricity \\
semi\_amplitude\_primary & Semi-amplitude of the radial velocity curve for the primary component (km/s) \\
semi\_amplitude\_secondary & Semi-amplitude of the radial velocity curve for the secondary component (km/s) \\
\hline
\multicolumn{2}{c}{\textit{Stellar Parameters}} \\
\hline
m1 & Mass of the primary component (solar masses) \\
m2 & Mass of the secondary component (solar masses) \\
fluxratio & Flux ratio of the secondary to the primary in the G-band ($F_2/F_1$) \\
\hline
\end{tabular}
\begin{flushleft}
\small Note: For additional details, see the Gaia DR3 documentation: \url{https://gea.esac.esa.int/archive/documentation/GDR3/Gaia_archive/chap_datamodel/sec_dm_non--single_stars_tables/}
\end{flushleft}
\end{subtable}
\end{table*}
\subsection{Cross-match with Gaia Non-Single Star catalog}\label{app:nss}

In addition to our primary binary star catalog, we provide a supplementary dataset cross-matching our identified binary systems with the Gaia DR3 Non-Single Star (NSS) catalog. Table~\ref{tab:nss_params} summarizes the key NSS parameters included in this cross-matched catalog. 
This cross-matched catalog is particularly valuable for systems with both spectrophotometric signatures (identified through our neural network approach) and dynamical evidence (from the Gaia NSS solutions), allowing for more robust mass ratio determinations and complete orbital characterization.

\end{appendix}

\end{document}